# Quantum-Enhanced Adversarial Robustness in Artificial Intelligence

Jaydip Sen
*Praxis Business School, India*

## ABSTRACT

*Artificial Intelligence (AI) has achieved remarkable success across diverse application domains; however, its vulnerability to adversarial attacks poses significant challenges to reliability, security, and trustworthiness. Adversarial machine learning demonstrates that even highly accurate models can be manipulated through carefully crafted perturbations, raising serious concerns in safety-critical systems such as healthcare, finance, and autonomous technologies. In parallel, quantum computing has emerged as a transformative paradigm capable of addressing complex computational problems through principles such as superposition, entanglement, and quantum interference. The convergence of these fields has led to the emergence of Quantum Artificial Intelligence (QAI), which explores how quantum techniques can enhance learning efficiency, scalability, and robustness. This chapter provides a comprehensive overview of adversarial machine learning and existing defense strategies, followed by an accessible introduction to quantum computing and quantum machine learning models. It further presents conceptual frameworks for quantum-enhanced adversarial robustness, emphasizing quantum optimization, feature mapping, and hybrid quantum–classical architectures. Practical applications, key challenges, and future research directions are also discussed to support the development of secure and trustworthy AI systems.*

**Keywords:** Quantum Artificial Intelligence, Quantum Machine Learning, Adversarial Machine Learning, Hybrid Quantum-Classical Learning, Secure Artificial Intelligence, Quantum Optimization Algorithms, Quantum-Enhanced Learning Models, AI Security and Resilience.

## INTRODUCTION

Artificial Intelligence (AI) has emerged as one of the most transformative technologies of the twenty-first century, fundamentally reshaping how information is processed, decisions are made, and services are delivered across digital ecosystems. Over the past decade, advances in machine learning algorithms, large-scale data availability, and high-performance computing infrastructure have enabled AI systems to achieve remarkable success in complex tasks such as image recognition, natural language understanding, predictive analytics, autonomous navigation, and intelligent recommendation systems. These developments have accelerated the deployment of AI technologies across diverse sectors including healthcare, finance, transportation, manufacturing, and intelligent infrastructure (Biamonte et al., 2017; Dunjko & Briegel, 2018; Devadas & Sowmya, 2025; Melnikov et al., 2023; Alexeev et al., 2025). AI-driven systems now support critical decision-making processes in medical diagnosis, financial risk assessment, industrial automation, fraud detection, and urban planning, thereby playing an increasingly important role in modern socio-economic systems. As summarized in Table 1, AI has evolved through multiple paradigms, each introducing new capabilities while also exposing new vulnerabilities.

*Table 1. Evolution of Artificial Intelligence and Emerging Paradigms*

| Era | Key Developments | Technologies | Limitations |
|---|---|---|---|
| Early AI (1950s-1990s) | Rule-based systems, symbolic AI | Expert systems | Lacks scalability |
| Machine Learning Era (2000-2015) | Statistical Learning, SVMs | Classical ML models | Feature engineering dependency |
| Deep Learning Era (2015-2022) | CNNs, RNNs, Transformers | Large neural networks | Vulnerability to adversarial attacks |
| Modern AI (2022-Present) | Generative AI, XAI, Agentic AI | LLMs, multimodal models | Security, robustness, trust issues |
| Emerging Paradigm | Quantum AI, Hybrid AI | QML, VQAs | Hardware and theoretical limitations |

Recent developments in AI have further accelerated this transformation. The emergence of Generative AI models capable of producing high-quality text, images, and multimedia content has demonstrated the potential of deep learning architectures to emulate complex aspects of human creativity and reasoning. At the same time, the growing emphasis on Explainable Artificial Intelligence (XAI) has attempted to address concerns regarding transparency and interpretability in machine learning models, particularly in domains where algorithmic decisions must be trusted and understood by human stakeholders (Hane et al., 2023; Devadas & Sowmya, 2025; Lu et al., 2024). Additionally, the development of agentic AI systems, which can autonomously perform tasks and interact with dynamic environments, has expanded the scope of AI applications to include adaptive systems capable of multi-step reasoning and decision-making. Together, these developments have significantly broadened the capabilities of AI technologies and strengthened their integration into real-world operational environments.

Despite these remarkable achievements, the increasing reliance on AI technologies has also raised critical concerns regarding the reliability, robustness, and security of machine learning systems. Modern AI models, particularly deep neural networks, are known to exhibit significant vulnerabilities that can be systematically exploited adversarial actors (Biggio & Roli, 2018; Chakraborty et al., 2021; Han et al., 2023; West et al., 2023b; Wendlinger et al., 2024; Sen et al., 2023). One of the most widely studied threats in this context is the phenomenon of adversarial attacks, where carefully crafted perturbations are introduced into input data in order to manipulate the output of machine learning models (Goodfellow et al., 2015; Szegedy et al., 2014; Biggio & Roli, 2018; Han et al., 2023; Wendlinger et al., 2024). These perturbations are typically small and often imperceptible to human observers, yet they can cause AI systems to produce incorrect predictions with high confidence (Goodfellow et al., 2015; Madry et al., 2018; Szegedy et al., 2014). Research has demonstrated that widely used models for image classification, speech recognition, and natural language processing can be fooled by such adversarial inputs, raising significant concerns about the reliability of AI systems deployed in critical environments.

Adversarial attacks can take many forms depending on the attacker's knowledge and capabilities. Gradient-based methods such as the Fast Gradient Sign Method (FGSM) (Goodfellow et al., 2015) and Projected Gradient Descent (PGD) (Madry et al., 2018) exploit the gradients of machine learning models to identify directions in which small perturbations can cause large changes in predictions. Optimization-based attacks and black-box attacks further demonstrate that adversarial vulnerabilities are not limited to specific models but can generalize across architectures due to the transferability property of adversarial examples across different models and architectures (Papernot et al., 2016; Biggio & Roli, 2018; Chakraborty et al., 2021; Sen et al., 2023). These findings indicate that adversarial threats are deeply rooted in the structural properties of machine learning models rather than being isolated anomalies.

The implications of adversarial attacks extend far beyond academic experimentations and pose serious risks in safety-critical applications. For example, adversarial perturbations may cause autonomous systems to misinterpret environmental signals, medical diagnostic systems to generate incorrect predictions, or financial fraud detection algorithms to overlook malicious transactions. In cybersecurity contexts, adversarial manipulation of machine learning models can undermine intrusion detection systems and malware classification tools. These vulnerabilities threaten the integrity and trustworthiness of AI-driven decision systems and highlight the need for more robust learning frameworks capable of operating securely in adversarial environments (Biggio & Roli, 2018; Chakraborty et al., 2021; Han et al., 2023; Sen et al., 2023).

In response to these concerns, researchers have proposed numerous defense strategies aimed at improving the robustness of machine learning models. Among the most widely studies approaches is adversarial training, which incorporates adversarial examples into the training process in order to expose models to potential attack patterns during learning (Madry et al., 2018). Other techniques include defensive distillation (Papernot et al., 2016), gradient masking (Biggio & Roli, 2018; Chakraborty et al., 2021), input preprocessing (Chakraborty et al., 2021), and robust optimization frameworks (Madry et al., 2018; Chakraborty et al., 2021). Although these methods have demonstrated partial success in mitigating certain attack types, they often suffer from significant limitations. Some defenses incur high computational costs, while others fail to generalize against adaptive adversarial attacks. These limitations suggest that fundamentally new approaches may be required to develop AI systems that are truly resilient against adversarial manipulation.

While researchers continue to investigate new defense mechanisms within classical computing paradigms, another technological revolution is unfolding in the field of quantum computing. Quantum computing represents a fundamentally different computational framework that leverages the principles of quantum mechanics to perform information processing tasks. Unlike classical bits, which can exist only in binary states of 0 or 1, quantum bits, i.e., qubits, can exist in superpositions of multiple states simultaneously. Additionally, quantum systems exhibit phenomena such as entanglement and quantum interference, which enable complex correlations between qubits and allow quantum algorithms to explore computational spaces in ways that are not possible for classical computers (Nielsen & Chuang, 2010; Preskill, 2018; Bharti et al., 2022).

Over the past decade, substantial progress has been made in both the hardware and software aspects of quantum computing. Advances in quantum hardware and algorithm design have opened new opportunities for solving complex problems in optimization, simulation, and machine learning (Bharti et al., 2022; Alexeev et al., 2025; Liu et al., 2024; Iovane, 2025). At the same time, researchers have developed a growing set of quantum algorithms that demonstrate potential advantages over classical approaches in specific problem domains. These developments have laid the foundation for integrating quantum computing techniques with artificial intelligence.

The convergence of quantum computing and artificial intelligence has given rise to an emerging interdisciplinary research field commonly referred to as Quantum Artificial Intelligence (QAI) or Quantum Machine Learning (QML) (Biamonte et al., 2017; Dunjko & Briegel, 2018, Arunachalam, S., & de Wolf, 2017; Devadas & Sowmya, 2025; Lu et al., 2024; Melnikov et al, 2023). This field investigates how quantum computational techniques can enhance machine learning algorithms and data analysis methods. Several quantum machine learning frameworks have been proposed, including variational quantum circuits (Biamonte et al., 2017; Benedetti et al., 2019; Cerezo  et al., 2021), quantum kernel methods (Havlíček et al., 2019; Schuld & Petruccione, 2021; Blank et al., 2020), and hybrid quantum-classical learning architectures (Cerezo et al., 2021; Abbas et al., 2021). These approaches aim to leverage quantum properties to improve learning performance, particularly in high-dimensional and complex data environments.

One of the most promising aspects of quantum computing for AI applications lies in its potential ability to handle complex optimization and high-dimensional search problems more efficiently than classical algorithms. Quantum optimization techniques such as the Quantum Approximate Optimization Algorithm (QAOA) (Farhi et al., 2022) and quantum annealing methods (Neukart et al., 2017; Pomeroy et al, 2025; Date et al., 2019; Kim et al., 2025; Heidari et al., 2024; Yulianti et al., 2023l; Salloum et al., 2024; Kotsuki et al., 2024) have been proposed as potential tools for exploring large solution spaces and identifying optimal or near-optimal solutions. These methods may offer advantages in training machine learning models, selecting features, and solving combinatorial optimization problems that are computationally intensive for classical systems (Farhi et al., 2022; Iovane, 2025; Kim et al, 2025).

The potential synergy between quantum computing and machine learning has motivated researchers to explore whether quantum techniques could address some of the fundamental challenges facing modern AI systems. One such challenge is adversarial robustness. Since adversarial attacks exploit weaknesses in optimization processes and feature representations, quantum computational techniques may provide new ways to enhance robustness. For example, quantum optimization algorithms may enable more effective adversarial training, while quantum feature mapping techniques may improve the separability of data in high-dimensional feature spaces, reducing susceptibility to adversarial perturbations (West et al., 2023a; West et al., 2023b; Gong et al., 2024; Dowling et al., 2026; Huang et al., 2023; Nowmi et al., 2025).

As illustrated in Figure 1, the evolution of AI has progressed through multiple paradigms, each contributing new capabilities while introducing new challenges.

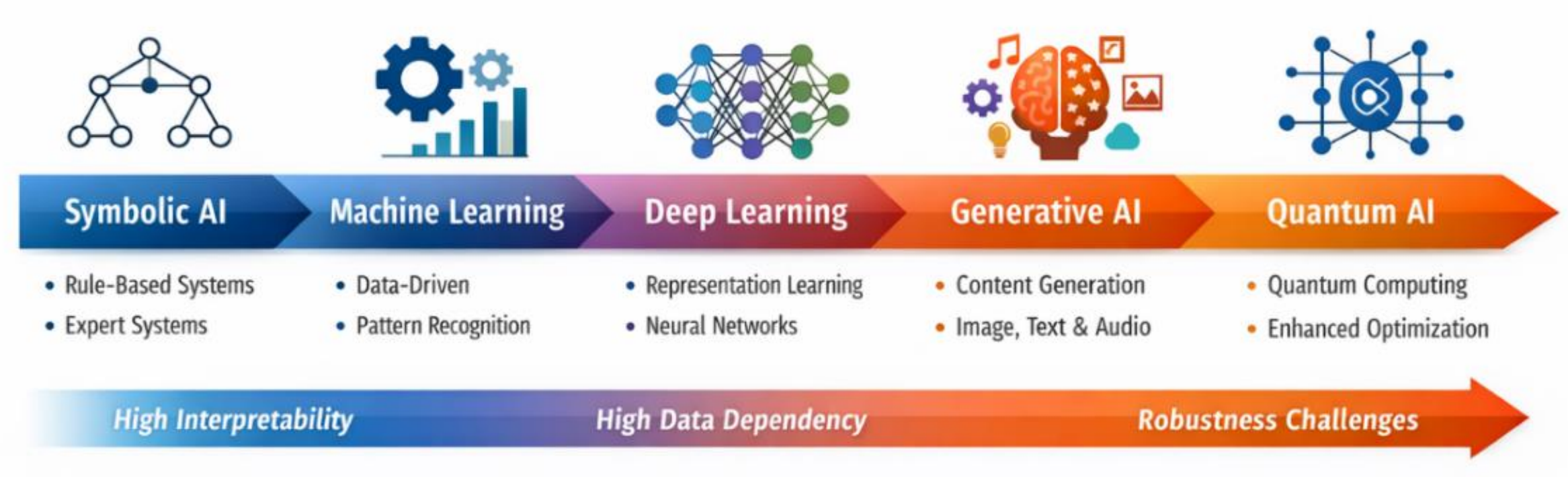

*Figure 1. Evolution of Artificial Intelligence and Emerging Paradigms*

Another promising direction involves the development of hybrid quantum-classical learning architectures that combine classical neural networks with parameterized quantum circuits (Benedetti et al., 2019; Cerezo et al., 2021; Abbas et al., 2021). These architectures allow quantum circuits to perform specialized transformations or probabilistic modeling tasks, while classical components handle large-scale data processing. Such hybrid approaches are particularly relevant in the current era of noisy intermediate-scale quantum (NISQ) devices, where fully quantum machine learning systems remain impractical, but hybrid solutions can still provide meaningful benefits.

In addition to improving model robustness, quantum computing may also contribute to the detection of adversarial activity. Quantum probabilistic models and anomaly detection techniques could potentially identify unusual patterns in data that indicate adversarial manipulation (Devadas & Sowmya, 2025; Melnikov et al., 2023; Liu et al., 2024). By analyzing data distributions in quantum feature spaces, AI systems may be able to detect subtle anomalies that are difficult to capture using classical methods. Although these approaches are still in early stages of development, they represent an important direction for future research.

Despite its promise, the integration of quantum computing with AI security faces significant challenges. Current quantum hardware is limited by noise, decoherence, and scalability constraints, which restrict the complexity of quantum algorithms that can be implemented in practice (Preskill, 2018; Bharti et al., 2022).

Furthermore, the theoretical foundations of quantum-enhanced adversarial defense mechanisms are still evolving, and many proposed approaches require further validation through both theoretical analysis and experimental studies. Bridging the gap between quantum computing and AI research communities also remains an important challenge that must be addressed to fully realize the potential of Quantum Artificial Intelligence (Preskill, 2018; Bharti et al., 2022, Alexeev et al., 2025; Liu et al., 2024).

The remainder of this chapter is organized as follows. The next section provides a comprehensive overview of adversarial machine learning, including common attack models, threat taxonomies, and existing defense strategies, along with their limitations. This is followed by a discussion of the fundamental concepts of quantum computing, including key ideas such as qubits, quantum gates, quantum circuits, and key quantum algorithms relevant to machine learning. Subsequently, the chapter examines the foundations of Quantum Artificial Intelligence, with a focus on quantum machine learning models, variational quantum circuits, and hybrid quantum–classical architectures. The discussion then introduces conceptual frameworks for quantum-enhanced adversarial robustness, highlighting how quantum optimization, feature mapping, and hybrid learning models can strengthen the security and reliability of AI systems. The chapter further explores practical applications of quantum-enhanced robust AI across domains such as healthcare, finance, transportation, and cybersecurity. This is followed by a discission on key challenges, limitations, and open research problems in the field. The chapter concludes by summarizing key insights and outlining future research directions in quantum-enhanced secure AI systems.

## ADVERSARIAL MACHINE LEARNING: ATTACKS AND DEFENSES

Adversarial machine learning has emerged as a critically important research domain within the broader field of artificial intelligence, driven by the rapid deployment of machine learning systems in real-world, high-impact applications. This area of study focuses on understanding the inherent vulnerabilities of machine learning models and developing systematic approaches to both exploit and defend against these weaknesses. As modern AI systems increasingly influence decision-making in domains such as healthcare diagnostics, financial systems, intelligent transportation, cybersecurity, and critical infrastructure, ensuring their robustness and security has become a fundamental requirement rather than a secondary concern (Biggio & Roli, 2018; Chakraborty et al., 2021; Han et al., 2023).

At its core, adversarial machine learning investigates how intelligent systems behave under *malicious or strategically manipulated inputs.* Unlike traditional machine learning settings, where data is assumed to be *independently and identically distributed* (i.i.d.) and free from manipulation, adversarial settings consider the presence of intelligent attackers who actively attempt to deceive models. These attackers exploit weaknesses in model architectures, training procedures, or data representations to craft adversarial examples—inputs that are intentionally designed to cause incorrect predictions while remaining indistinguishable from legitimate data to human observers (Goodfellow et al., 2015; Szegedy et al., 2014). This shift from passive to adversarial environments introduces a new dimension of complexity in the design and evaluation of machine learning systems. Figure 2 presents a general framework illustrating how adversarial perturbations manipulate model predictions.

Adversarial attacks can occur at multiple stages of the machine learning lifecycle, leading to a diverse range of threat models. *Evasion attacks,* for example, occur during the inference phase, where attackers manipulate input data to mislead trained models without altering the training process. In contrast, *poisoning attacks* target the training phase by injecting malicious data into the training dataset, thereby corrupting the learned model. Additionally, *model extraction and inversion attacks* aim to infer sensitive information about the model or its training data, raising significant privacy concerns (Biggio & Roli, 2018; Chakraborty et al., 2021). These varied attack strategies demonstrate that adversarial threats are not limited to a single point of vulnerability but can affect the entire machine learning pipeline.

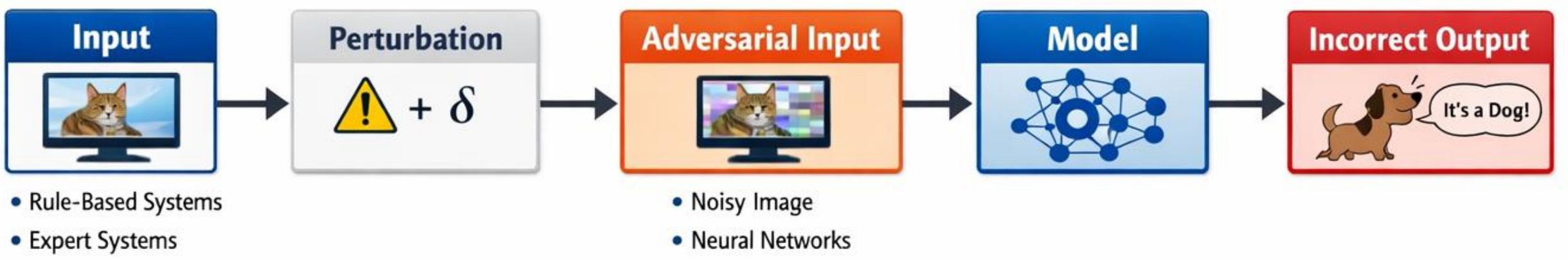

*Figure 2. General Framework of Adversarial Attacks on Machine Learning Models*

The growing sophistication of adversarial attack techniques has been accompanied by the development of a wide range of *defense mechanisms* aimed at enhancing model robustness. These defenses seek to either improve the intrinsic resilience of machine learning models or detect and mitigate adversarial inputs before they can cause harm. Approaches such as adversarial training, robust optimization, input preprocessing, and anomaly detection have shown varying degrees of effectiveness in countering specific attack types (Madry et al., 2018; Papernot et al., 2016; Han et al., 2023). However, many defense strategies suffer from limitations, including high computational costs, reduced performance on clean data, and vulnerability to adaptive attacks that evolve in response to deployed defenses.

A key challenge in adversarial machine learning is the *dynamic and evolving nature of the threat landscape.* Attackers continuously develop new strategies to bypass existing defenses, leading to an ongoing arms race between attack and defense techniques. This adversarial interplay highlights the need for defense mechanisms that are not only effective against known attacks but also generalizable to unseen and adaptive threats. Moreover, the transferability of adversarial examples across different models further complicates the design of robust systems, as vulnerabilities may persist even when models are modified or retrained (Papernot et al., 2016; Biggio & Roli, 2018).

Another important dimension of adversarial machine learning is its connection to broader concerns related to *trustworthiness, fairness, and accountability in AI systems*. Vulnerabilities to adversarial manipulation can undermine user trust, compromise system integrity, and lead to unintended consequences in critical applications. For instance, adversarial attacks on medical diagnostic systems could result in incorrect treatment decisions, while attacks on autonomous systems could pose safety risks. These implications underscore the importance of integrating robustness and security considerations into the design of AI systems from the outset, rather than treating them as afterthoughts.

Given these challenges, adversarial machine learning is increasingly viewed not only as a security problem but also as a *fundamental limitation of current machine learning paradigms*. Addressing these limitations requires a deeper understanding of the theoretical properties of learning algorithms, data distributions, and optimization processes. Emerging research directions, including robust statistics, certified defenses, and novel computational paradigms such as quantum-enhanced learning, offer promising avenues for developing more resilient AI systems.

In summary, adversarial machine learning represents a crucial frontier in AI research, bridging the gap between machine learning, cybersecurity, and system reliability. By systematically studying both attack strategies and defense mechanisms, this field aims to build intelligent systems that can operate safely and reliably in adversarial environments. As AI continues to permeate critical aspects of society, advancements in adversarial robustness will play a central role in ensuring the long-term sustainability and trustworthiness of intelligent technologies.

## Taxonomy of Adversarial Attacks

A comprehensive understanding of adversarial machine learning necessitates the development of a well-structured taxonomy of adversarial attack methods. Such a taxonomy enables researchers and practitioners to systematically analyze different threat models, evaluate vulnerabilities, and design effective defense mechanisms. Existing literature broadly categorizes adversarial attacks along multiple dimensions, including the attacker's knowledge of the target model, the attacker's objective, the stage of the machine learning pipeline being targeted, and the techniques used to generate adversarial perturbations (Biggio & Roli, 2018; Chakraborty et al., 2021). This multi-dimensional classification provides a holistic framework for understanding the diverse landscape of adversarial threats. The classification of adversarial attacks is depicted in Figure 3.

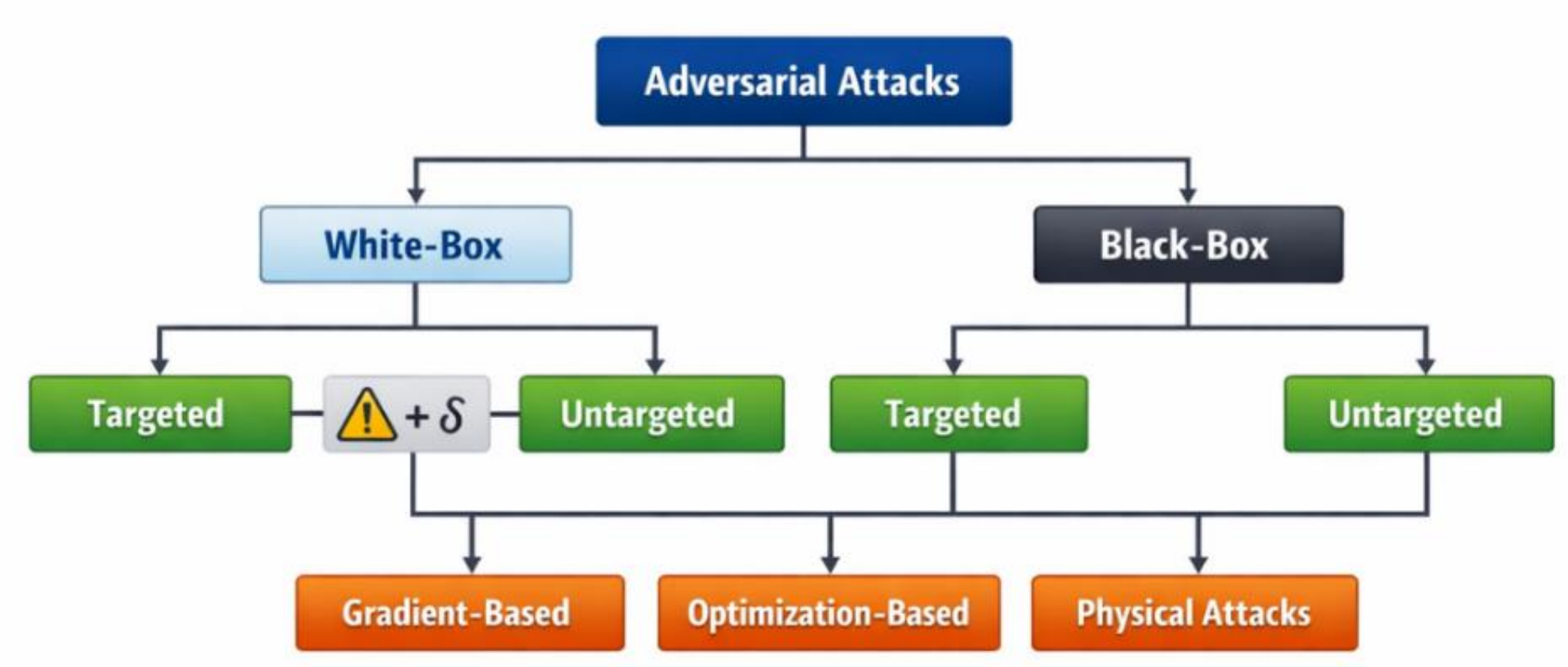

*Figure 3. Taxonomy of Adversarial Attacks*

From the perspective of *attacker knowledge,* adversarial attacks are typically divided into *white-box, gray-box,* and *black-box attacks.* In white-box settings, the attacker has complete access to the model architecture, parameters, and training data, enabling the use of highly effective gradient-based methods. In contrast, black-box attacks assume no internal knowledge of the model, requiring attackers to rely on input-output queries or surrogate models. Gray-box attacks represent an intermediate scenario where partial information is available. This classification is crucial because it reflects real-world conditions where attackers often operate under varying levels of information access.

Another important dimension is the *attacker's objective,* which determines the nature of the adversarial manipulation. Attacks may be *targeted,* where the goal is to force the model to produce a specific incorrect output, or *untargeted,* where any incorrect prediction is sufficient. Additionally, adversarial attacks can be categorized based on whether they aim to degrade overall model performance, manipulate specific predictions, or extract sensitive information from the model. These objectives influence the design and effectiveness of attack strategies.

Adversarial attacks can also be classified based on the *stage of the machine learning lifecycle* they target. *Evasion attacks* occur during the inference phase, where adversarial inputs are crafted to deceive a trained model without altering its parameters. *Poisoning attacks,* on the other hand, target the training phase by injecting malicious or misleading data into the training dataset, thereby corrupting the learned model. A related class of attacks includes *backdoor* or *Trojan attacks,* where hidden triggers are embedded in the training data to cause specific misclassifications when activated. This lifecycle-based classification highlights that vulnerabilities can arise at multiple stages of model development and deployment.

***Gradient-Based Attacks:*** Gradient-based attacks represent one of the most extensively studied and practically effective classes of adversarial techniques. These methods leverage the gradients of the model's loss function with respect to input features to identify directions in which small perturbations can significantly increase prediction error. The fundamental idea is to exploit the sensitivity of neural networks to input variations by moving the input data in the direction that maximizes the loss.

One of the earliest and most influential methods in this category is the *Fast Gradient Sign Method* (FGSM) (Goodfellow et al., 2015), which generates adversarial examples by applying a single-step perturbation in the direction of the gradient sign. Despite its simplicity, FGSM is highly effective and computationally efficient, making it a widely used baseline for evaluating model robustness. Building on this approach, *Projected Gradient Descent* (PGD) (Madry et al., 2018) introduces an iterative procedure that applies multiple small perturbations while projecting the perturbed input back into a constrained region. PGD is often considered one of the strongest first-order adversarial attacks and is commonly used as a benchmark for robust training.

More advanced gradient-based methods include *iterative gradient attacks, momentum-based attacks,* and adaptive techniques that exploit specific model characteristics. These methods improve attack success rates by stabilizing optimization trajectories and escaping local minima. Gradient-based attacks are particularly effective in white-box settings, where full access to model gradients is available. However, their effectiveness can diminish in black-box scenarios unless combined with transferability or gradient estimation techniques.

***Optimization-Based Attacks:*** Optimization-based attacks formulate the generation of adversarial examples as a constrained optimization problem. The objective is to find the smallest possible perturbation that causes a model to misclassify an input while maintaining perceptual similarity to the original data. This formulation provides a more principled approach to adversarial example generation compared to simple gradient-based methods.

The *Carlini–Wagner (C&W) attack* is widely regarded as one of the strongest optimization-based white-box adversarial attacks due to its ability to produce low-distortion adversarial examples that evade detection mechanisms (Carlini & Wagner, 2017a; Carlini & Wagner, 2017b). The attack uses carefully designed objective functions and optimization strategies to produce minimal yet highly effective perturbations. Unlike simpler methods, it explicitly balances perturbation magnitude and attack success, making it particularly powerful against defenses such as defensive distillation.

Although optimization-based attacks often achieve superior performance in terms of stealth and effectiveness, they come with increased computational costs due to iterative optimization procedures. This trade-off between computational complexity and attack strength makes them particularly relevant in scenarios where attackers are willing to invest additional resources to achieve high success rates.

***Transfer-Based and Black-Box Attacks:*** In many real-world applications, attackers do not have direct access to model internals, necessitating the use of *black-box attack strategies.* One of the most important phenomena enabling such attacks is the *transferability of adversarial examples,* where inputs crafted to deceive one model can also mislead other models with similar decision boundaries (Papernot et al., 2016; Biggio & Roli, 2018; Chakraborty et al., 2021). This property allows attackers to train surrogate models and generate adversarial examples that generalize to the target model.

*Transfer-based attacks* leverage this phenomenon by first creating adversarial examples using a substitute model and then applying them to the target model. Despite differences in architecture or training data, many models exhibit similar vulnerabilities, making transfer-based attacks surprisingly effective.

Another class of black-box attacks involves *query-based methods,* where attackers interact with the target model by submitting inputs and observing outputs. By analyzing these responses, attackers can estimate gradients or approximate decision boundaries without explicit access to model parameters. Techniques such as finite-difference gradient estimation and score-based optimization enable the construction of adversarial examples even under strict access limitations.

These attacks highlight the practical feasibility of adversarial manipulation in real-world systems, where direct access to model internals is often restricted. They also emphasize the importance of designing defenses that are robust not only in white-box settings but also against black-box and adaptive adversaries.

***Physical and Real-World Attacks:*** While early research in adversarial machine learning focused primarily on digital environments, recent studies have demonstrated that adversarial attacks can be extended to *physical and real-world settings,* significantly increasing their practical implications. In these scenarios, adversarial perturbations are embedded into physical objects or signals in a way that remains effective even after environmental transformations such as lighting changes, noise, and viewpoint variations.

One notable example is the manipulation of *traffic signs,* where carefully designed perturbations can cause autonomous driving systems to misclassify stop signs or speed limit indicators. Similarly, adversarial audio perturbations have been used to deceive *speech recognition systems,* enabling hidden commands that are imperceptible to human listeners. These attacks demonstrate that adversarial vulnerabilities are not limited to digital inputs but can manifest in real-world interactions between AI systems and their environments.

The existence of physical adversarial attacks underscores the urgency of addressing robustness in safety-critical applications. Unlike digital attacks, which may be contained within computational systems, physical attacks can have direct consequences on human safety and infrastructure. As a result, developing defenses that remain effective under real-world conditions is an important and ongoing research challenge.

***Summary:*** The taxonomy of adversarial attacks reveals the breadth and complexity of threats facing modern machine learning systems. From gradient-based and optimization-driven methods to black-box and physical attacks, adversarial techniques exploit fundamental weaknesses in how models learn and generalize. This diversity of attack strategies highlights the need for comprehensive and multi-layered defense mechanisms that can address vulnerabilities across different threat models and operational contexts.

Understanding this taxonomy is essential for advancing research in adversarial robustness and provides a foundation for exploring novel approaches, including *quantum-enhanced methods,* which may offer new capabilities for handling complex optimization landscapes and high-dimensional feature spaces. As adversarial machine learning continues to evolve, a systematic and theoretically grounded approach to both attacks and defenses will be critical for building secure and trustworthy AI systems.

A structured taxonomy of adversarial attacks is presented in Table 2, highlighting their classification based on attacker knowledge, objectives, and methodologies.

*Table 2. Taxonomy of Adversarial Attacks in Machine Learning*

| Category | Sub-Type | Description | Example Methods |
| --- | --- | --- | --- |
| Knowledge-Based | White box | Full model access | FGSM, PGD |
| Objective-Based | Targeted | Force specific output | CW attack |
| Method-Based | Gradient-based | Uses gradients | PGD |
| | Optimization-based | Solves optimization problem | CW |
| | Query-based | Model probing | Boundary attack |
| Environment-Based | Digital | Pixel/text perturbation | Image attacks |
| | Physical | Real-world manipulation | Adversarial signs |

## Vulnerabilities in Deep Learning Models

The susceptibility of deep learning models to adversarial attacks is a well-documented and fundamental challenge in modern machine learning. These vulnerabilities arise from a combination of factors related to data representation, model architecture, the training dynamics, and theoretical limitations of learning algorithms. Understanding these underlying causes is essential for developing robust and secure AI systems capable of operating reliably in adversarial environments.

One of the primary reasons for adversarial vulnerability is the *high dimensionality of input data* commonly encountered in deep learning applications. In domains such as image processing, natural language processing, and speech recognition, input data often reside in extremely high-dimensional spaces. In such spaces, even small perturbations to input features can lead to disproportionately large changes in model outputs. This phenomenon is partly explained by the concentration of measure in high-dimensional geometry, where data points are sparsely distributed, and decision boundaries can be easily crossed with minimal perturbations (Szegedy et al., 2014; Biggio & Roli, 2018). As a result, adversarial examples can be constructed by introducing imperceptible changes to input data that push it across the decision boundary, leading to incorrect predictions with high confidence. Figure 4 demonstrates how imperceptible perturbations can lead to incorrect predictions.

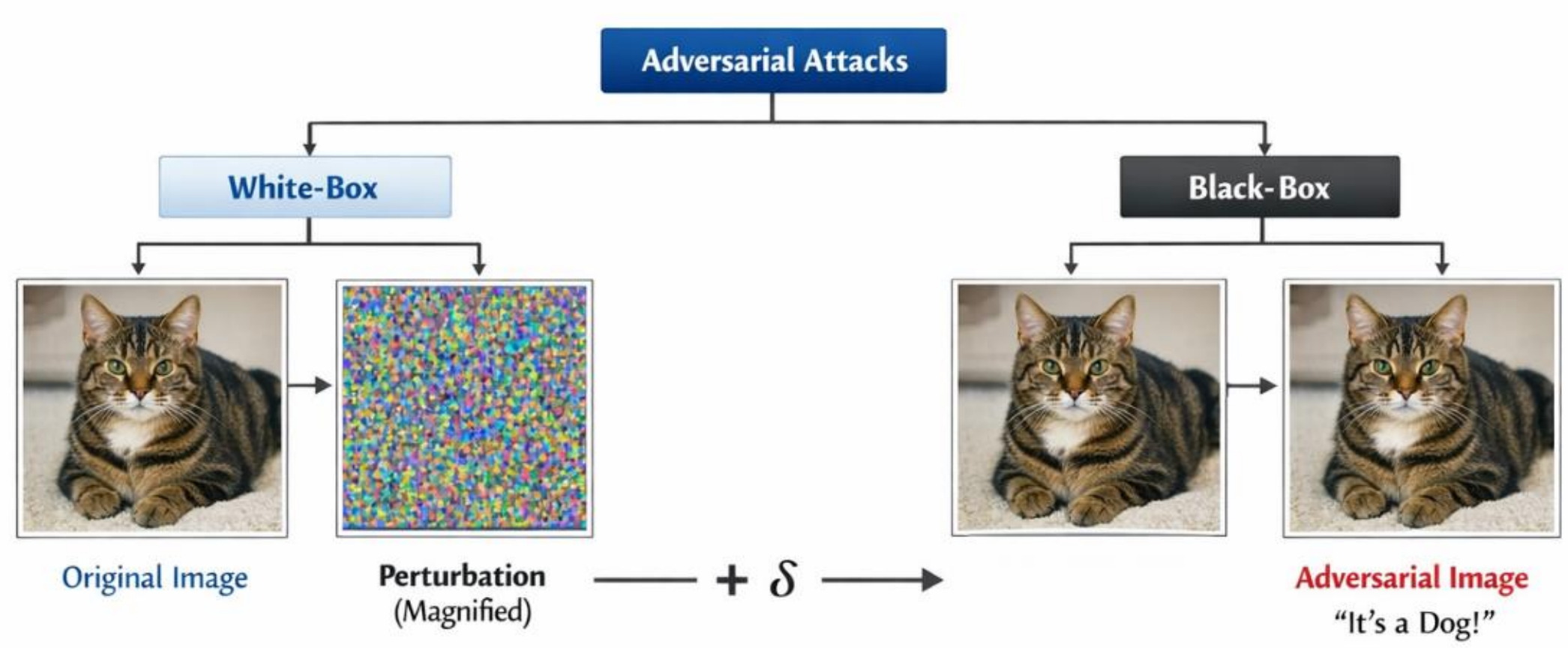

*Figure 4. Illustration of Adversarial Perturbation in Image Classification*

Closely related to this issue is the nature of *complex, non-linear decision boundaries* learned by deep neural networks. These models are designed to capture intricate patterns in data through multiple layers of nonlinear transformations. While this capability enables high predictive performance, it also results in highly irregular and sometimes fragile decision surfaces that may not generalize well beyond the training distribution. In many cases, the learned decision boundaries are overly sensitive to small perturbations in certain regions of the input space, making them susceptible to adversarial manipulation (Goodfellow et al., 2015; Szegedy et al., 2014). This sensitivity is further exacerbated by the fact that deep models often rely on high-frequency or non-robust features that are not aligned with human perception.

Another significant contributing factor is the issue of *overfitting and reliance on spurious correlations*. Deep learning models, particularly when trained on large but imperfect datasets, may learn patterns that are specific to the training data but do not generalize to unseen data. Instead of capturing true underlying relationships, models may exploit statistical regularities or noise present in the dataset. Adversarial attacks can exploit these weaknesses by introducing perturbations that align with these non-robust features, thereby misleading the model without altering the semantic content of the input (Biggio & Roli, 2018; Chakraborty et al., 2021; Han et al., 2023). This highlights a fundamental gap between model learning and human perception, where models fail to capture semantically meaningful features in a robust manner.

The *lack of interpretability and transparency* in deep learning models further complicates the understanding and mitigation of adversarial vulnerabilities. Deep neural networks are often regarded as "black-box" systems, where the internal decision-making process is difficult to interpret. This opacity makes it challenging to identify which features are being used for classification and how perturbations influence model predictions. Consequently, diagnosing vulnerabilities and designing effective defenses becomes significantly more difficult (Han et al., 2023). Efforts in explainable AI (XAI) aim to address this issue by providing insights into model behavior, but these approaches are still evolving and may not fully capture the complexities of adversarial interactions.

Another important factor contributing to adversarial susceptibility is the *linearity hypothesis* proposed by Goodfellow et al. (Goodfellow et al., 2015), which suggests that despite their non-linear architectures, many deep neural networks behave approximately linearly in high-dimensional spaces. This quasi-linear behavior enables attackers to exploit gradient information efficiently, generating adversarial perturbations that accumulate across dimensions to produce significant changes in output predictions. This insight has led to the development of powerful gradient-based attack methods and has deepened the understanding of why adversarial examples are so prevalent across different models and datasets.

Furthermore, the *transferability of adversarial examples* represents a critical challenge in adversarial machine learning. Adversarial inputs generated for one model often remain effective against other models, even when their architectures or training data differ. This property suggests that different models learn similar decision boundaries or feature representations, which can be exploited by attackers in black-box scenarios where model internals are not accessible (Papernot et al., 2015; Biggio & Roli, 2018; Chakraborty et al., 2021). Transferability significantly increases the practical feasibility of adversarial attacks and underscores the systemic nature of these vulnerabilities.

Biggio and Roli (Biggio & Roli, 2018) and Chakraborty et al. (Chakraborty et al., 2021) has emphasized that adversarial vulnerabilities are not isolated to specific architectures or training procedures but are inherent to a wide range of machine learning algorithms. These studies highlight those adversarial weaknesses stem from fundamental characteristics of statistical learning, including high-dimensional representations, empirical risk minimization, and the mismatch between training and testing distributions. More recent analyses (Han et al., 2023) further reinforce the idea that adversarial examples are deeply rooted in the geometry of data and the optimization processes used in training modern machine learning models.

Collectively, these findings indicate that improving the robustness of deep learning systems requires more than incremental modifications to existing models. Instead, it calls for a deeper understanding of the theoretical foundations of machine learning and the exploration of fundamentally new paradigms for learning and computation. In this context, emerging approaches such as robust optimization, probabilistic modeling, and quantum-enhanced learning frameworks offer promising directions for addressing the inherent limitations of classical machine learning systems and building more resilient AI architectures.

## Defense Mechanisms in Adversarial Machine Learning

In response to the growing prevalence and sophistication of adversarial attacks, the research community has proposed a wide range of defense mechanisms aimed at improving the robustness, reliability, and security of machine learning systems. These defense strategies can broadly be categorized into two main classes: (i) methods that enhance the intrinsic robustness of models during training, and (ii) techniques that attempt to detect or mitigate adversarial inputs at inference time. Despite substantial progress in this area, designing universally effective defenses remains a challenging task due to the adaptive nature of adversaries and the evolving landscape of attack methodologies (Biggio & Roli, 2018; Chakraborty et al., 2021; Han et al., 2023).

A fundamental challenge in adversarial defense lies in the trade-off between model robustness, computational efficiency, and generalization performance. Many defense mechanisms that demonstrate strong robustness against specific attack types often fail when confronted with adaptive or previously unseen adversarial strategies. Consequently, current research emphasizes the need for principled and theoretically grounded approaches that can provide reliable guarantees of robustness under diverse threat models.

***Adversarial Training:*** Adversarial training is widely regarded as one of the most effective and principled defense strategies against adversarial attacks. The core idea behind this approach is to augment the training dataset with adversarially perturbed examples, thereby enabling the model to learn representations that are robust to malicious perturbations. Formally, adversarial training can be framed as a min–max optimization problem in which the model is trained to minimize the worst-case loss under bounded perturbations of the input data (Madry et al., 2018).

This approach has demonstrated significant success in improving robustness against gradient-based attacks such as Fast Gradient Sign Method (FGSM) and Projected Gradient Descent (PGD), particularly when the adversarial examples are generated using strong iterative attack methods during training (Goodfellow et al., 2015; Madry et al., 2018). By exposing the model to adversarial samples during training, the learned decision boundaries tend to become smoother and more resilient to small perturbations in the input space.

However, adversarial training is not without limitations. One of its primary challenges is the substantial computational overhead associated with generating adversarial examples on-the-fly during training, especially for large-scale datasets and deep neural networks. Additionally, adversarially trained models often exhibit reduced accuracy on clean (non-adversarial) data, reflecting a trade-off between robustness and standard generalization performance. Another limitation is that adversarial training tends to be attack-specific; models trained against a particular class of attacks may remain vulnerable to other types of adversarial strategies, particularly black-box or transfer-based attacks (Biggio & Roli, 2018; Chakraborty et al., 2021).

Recent research has attempted to address these limitations by developing more efficient adversarial training methods, including curriculum-based adversarial training, ensemble adversarial training, and robust optimization-based approaches. Nevertheless, achieving scalable and generalizable adversarial training remains an open research problem.

***Defensive Distillation:*** Defensive distillation is another notable approach aimed at improving the robustness of machine learning models. Originally proposed by Papernot et al. (Papernot et al., 2015), this technique builds upon the concept of knowledge distillation, where a secondary model is trained using the soft output probabilities (logits) of a pre-trained network rather than hard class labels. The use of softened outputs, controlled by a temperature parameter, is intended to produce smoother decision boundaries and reduce the sensitivity of the model to small input perturbations.

The intuition behind defensive distillation is that by learning from probabilistic outputs that capture inter-class relationships, the model becomes less susceptible to abrupt changes in predictions caused by adversarial perturbations. Early studies suggested that this approach could significantly reduce the effectiveness of certain gradient-based attacks by diminishing the magnitude of gradients with respect to the input.

However, subsequent research has revealed critical weaknesses in defensive distillation. It has been shown that this method often relies on gradient obfuscation rather than true robustness, making it vulnerable to more sophisticated attack techniques that circumvent or approximate gradients (Biggio & Roli, 2018; Chakraborty et al., 2021). Adaptive adversaries can exploit these weaknesses by designing attacks that specifically target the defense mechanism, thereby restoring the effectiveness of adversarial perturbations. As a result, defensive distillation is no longer considered a reliable standalone defense, although it has contributed valuable insights into the importance of smooth decision boundaries in robust learning.

***Gradient Masking and Obfuscation:*** Gradient masking, also referred to as gradient obfuscation, encompasses a class of defense strategies that attempt to hide or distort gradient information in order to impede gradient-based adversarial attacks. Since many powerful attacks, including FGSM and PGD, rely on access to gradients for generating adversarial perturbations, obscuring this information can make such attacks more difficult to execute.

Techniques for gradient masking include introducing non-differentiable operations, stochastic components, or numerical instabilities into the model, thereby making gradient computation unreliable or misleading. While these methods may initially appear effective against standard attacks, they often provide only a false sense of security. Athalye et al. and subsequent studies have demonstrated that gradient masking defenses can be systematically bypassed using techniques such as gradient approximation, backward pass differentiation, or attack transferability (Athalye et al., 2018; Biggio & Roli, 2018; Chakraborty et al., 2021).

A key limitation of gradient masking is that it does not address the underlying vulnerability of the model but merely obscures the attack surface. Consequently, once the defense mechanism is understood, attackers can design more sophisticated strategies to circumvent it. For this reason, gradient masking is generally discouraged as a primary defense strategy and is instead viewed as an artifact to be avoided when designing robust models.

***Input Transformation and Preprocessing:*** Input transformation and preprocessing techniques represent another class of defenses that aim to mitigate adversarial attacks by modifying input data before it is processed by the model. The underlying assumption is that adversarial perturbations often exhibit specific statistical or structural properties that can be filtered or suppressed through appropriate transformations.

Common techniques in this category include feature squeezing, input denoising, image compression, random resizing, and data augmentation. Feature squeezing, for example, reduces the dimensionality or precision of input features in order to eliminate small perturbations, while denoising methods attempt to reconstruct clean inputs from corrupted data. Similarly, random transformations can introduce variability that disrupts the effectiveness of adversarial perturbations.

Although these approaches can provide some level of protection against certain types of attacks, their effectiveness is often limited. One major drawback is that input transformations may degrade the quality of legitimate data, leading to reduced model accuracy on clean inputs. Furthermore, adaptive adversaries can design attacks that are robust to these transformations, thereby bypassing the defense mechanism (Chakraborty et al., 2021). As a result, input preprocessing techniques are typically used in conjunction with other defense strategies rather than as standalone solutions.

A high-level overview of adversarial defense strategies is shown in Figure 5.

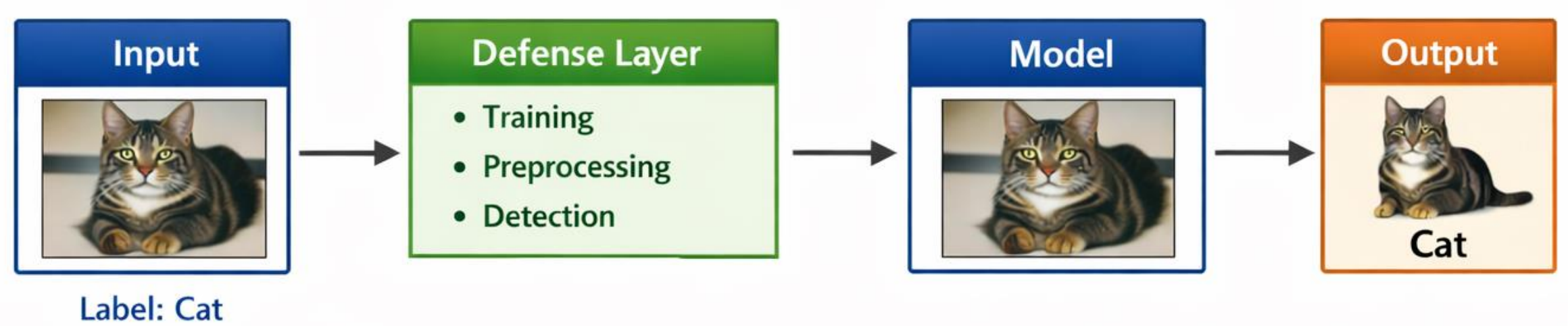

*Figure 5. Overview of Adversarial Defense Mechanisms*

***Robust Optimization:*** Robust optimization provides a theoretically grounded framework for enhancing the resilience of machine learning models against adversarial attacks. In this approach, the training objective is explicitly formulated to account for worst-case perturbations within a predefined uncertainty set. This leads to optimization problems that seek to minimize the maximum possible loss under adversarial conditions, thereby ensuring that the model performs reliably even in the presence of malicious inputs (Madry et al., 2018).

The robust optimization framework has strong connections to adversarial training, as both approaches aim to incorporate adversarial considerations into the learning process. However, robust optimization emphasizes formal guarantees and theoretical analysis, often leveraging tools from convex optimization, game theory, and statistical learning theory.

Despite its conceptual appeal, robust optimization faces several practical challenges. The computational complexity of solving min–max optimization problems can be prohibitive, particularly for deep neural networks with high-dimensional input spaces. Additionally, defining appropriate uncertainty sets that accurately capture real-world adversarial scenarios remains a non-trivial task. Overly restrictive assumptions may lead to limited robustness, while overly broad assumptions can result in overly conservative models with reduced performance on clean data.

Recent advancements have explored more efficient approximations and scalable algorithms for robust optimization, as well as hybrid approaches that combine robust training with other defense mechanisms. Nevertheless, achieving scalable, generalizable, and theoretically sound robustness remains an open challenge in adversarial machine learning (Biggio & Roli, 2018; Chakraborty et al., 2021; Han et al., 2023).

***Limitations of Existing Defense Mechanisms:*** Despite the diversity of defense mechanisms proposed in the literature, no single approach has proven to be universally effective against all types of adversarial attacks. Many defenses are vulnerable to adaptive adversaries who can tailor their strategies to exploit specific weaknesses in the defense mechanism. Moreover, the trade-offs between robustness, computational efficiency, and generalization performance continue to pose significant challenges.

These limitations highlight the need for fundamentally new paradigms for enhancing the security and robustness of AI systems. In this context, emerging computational frameworks such as quantum computing offer intriguing possibilities for developing novel defense mechanisms that go beyond the limitations of classical approaches. The following sections explore how quantum computational techniques can be leveraged to address these challenges and contribute to the development of more robust and secure AI systems.

Table 3 compares widely used defense mechanisms and highlights their strengths and limitations.

*Table 3. Comparison of Major Adversarial Defense Mechanisms*

| Defense Method | Principle | Strengths | Limitations |
|---|---|---|---|
| Adversarial Training | Train with adversarial samples | Strong empirical robustness | High computational cost |
| Defensive Distillation | Smooth decision boundaries | Reduces sensitivity | Breakable by adaptive attacks |
| Gradient Masking | Obscure gradients | Simple implementation | False security |
| Input Transformation | Preprocess inputs | Easy integration | Accuracy trade-off |
| Robust Optimization | Worst-case training | Theoretical foundation | Scalability issues |

## FUNDAMENTALS OF QUANTUM COMPUTING

Quantum computing represents a paradigm shift in computation, grounded in the principles of quantum mechanics. Unlike classical computing, which relies on deterministic binary states, quantum computing leverages probabilistic and wave-like behaviors of physical systems to process information. This enables fundamentally new computational capabilities, particularly for problems involving high-dimensional search, optimization, and simulation.

The theoretical foundations of quantum computing are well established (Neilsen & Chuang, 2010), and recent advances in quantum hardware and algorithm design have accelerated their practical development (Preskill, 2018; Bharti et al., 2022). These developments have positioned quantum computing as a promising technological foundation for next-generation intelligent systems.

This section provides a comprehensive overview of the fundamental concepts of quantum computing. The discussion covers quantum bits (qubits), quantum gates, quantum circuits, and key quantum algorithms, with an emphasis on their applicability to machine learning and data-driven systems.

## Quantum Bits and Quantum States

The fundamental unit quantum computation is the quantum bit (or qubit), which differs significantly from the classical bit. While a classical bit can exist in one of two states, either 0 or 1, a qubit can exist in a *superposition* of both states simultaneously. Mathematically the state of a qubit can be expressed as in (1)

$$|\psi\rangle = \alpha|0\rangle + \beta|1\rangle$$

In (1), $\boldsymbol{\alpha}$ and $\boldsymbol{\beta}$ are complex probability amplitudes such that $|\boldsymbol{\alpha}|^2 + |\boldsymbol{\beta}|^2 = \mathbf{1}$. This representation allows a qubit to encode a continuum of states, enabling quantum systems to process a vast amount of information in parallel (Nielsen & Chuang, 2010; Schuld & Petruccione, 2018).

A useful way to visualize a qubit is through Bloch sphere, a geometric representation in which any pure quantum state corresponds to point on the surface of a unit sphere (Nielsen & Chuang, 2010; Schuld & Petruccione, 2018; Dunjko & Briegel, 2018). This visualization highlights the continuous nature of quantum states and the rich structure of quantum state space compared to classical binary systems. The geometric representation of a qubit using the Bloch sphere is shown in Figure 6.

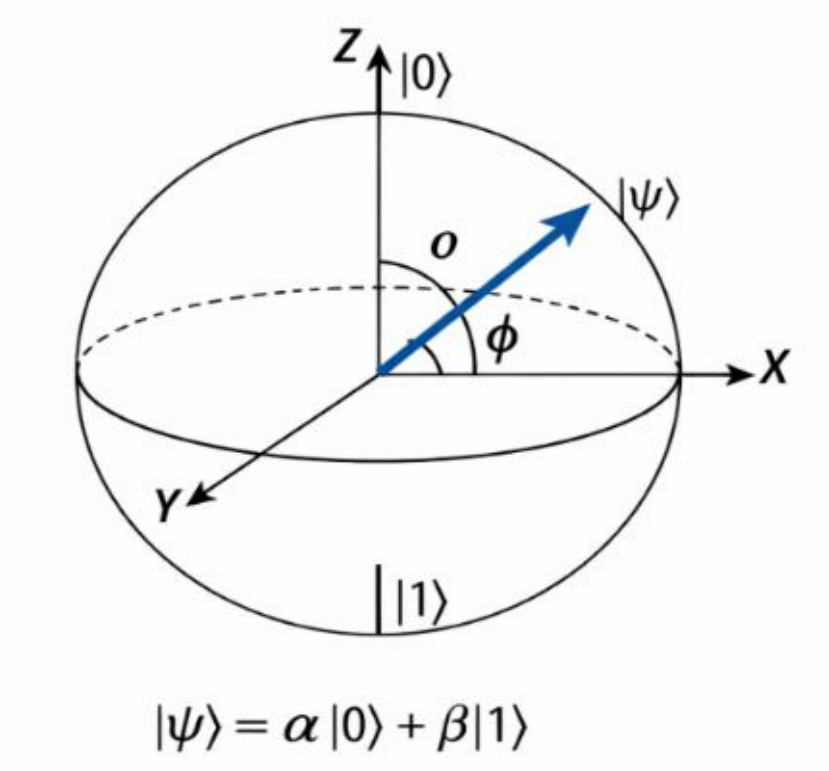

*Figure 6. Bloch Sphere Representation of a Qubit*

Another fundamental property of qubits is *entanglement*, a unique quantum phenomenon in which the state of one qubit becomes intrinsically correlated with the state of another, regardless of the distance between them. Entanglement enables quantum systems to represent complex correlations across multiple variables, which is particularly relevant for machine learning tasks involving high-dimensional data (Nielsen & Chuang, 2020; Preskill, 2018).

In addition to superposition and entanglement, quantum systems exhibit *quantum interference*, which allows probability amplitudes to combine constructively or destructively. This property is central to the design of quantum algorithms, enabling them to amplify correct solutions while suppressing incorrect ones.

## Quantum Gates and Operations

Quantum computation is performed through the application of *quantum gates*, which are unitary transformations that manipulate the state of qubits. Unlike classical logic gates, which are irreversible in many cases, quantum gates are inherently *reversible* and must preserve the total probability of the system (Nielsen & Chuang, 2020; Schuld & Petruccione, 2018).

Single-qubit gates are the simplest type of quantum operations. Common examples include the following:

- **Pauli-X gate:** Analogous to the classical NOT gate, it flips the computational basis states of a qubit, mapping $|\mathbf{0}\rangle$ to $|\mathbf{1}\rangle$ and vice versa.
- **Pauli-Y and Pauli-Z gates:** Represent rotations of a qubit state around the Y-axis and Z-axis of the Bloch sphere, respectively, thereby enabling controlled manipulation of quantum phase and amplitude.
- **Hadamard gate (H):** Creates quantum superposition by transforming a computational basis state into an equal linear combination of $|\mathbf{0}\rangle$ and $|\mathbf{1}\rangle$, thereby enabling parallel exploration of multiple states in quantum computation.
- **Phase gates:** Introduce controlled relative phase shifts between quantum states and modifies the phase of a qubit without altering its probability amplitudes. These gates are essential for enabling quantum interference effects.

Multi-qubit gates enable interactions between qubits and are essential for creating entanglement. The most commonly used multi-qubit gate is the *Controlled-NOT* (CNOT) gate, which flips the state of target qubit conditioned on the state of a control qubit. More complex gates, such as Toffoli and controlled-phase gates, enable the construction of sophisticated quantum circuits (Nielsen & Chuang, 2020; Cerezo et al., 2021).

Quantum gates can be mathematically represented as unitary matrices, and their sequential application defines the evolution of a quantum system. The design of quantum algorithms involves carefully orchestrating sequences of such gates to achieve a desired computational outcome (Nielsen & Chuang, 2020; Schuld & Petruccione, 2018).

## Quantum Circuits

A quantum circuit is a structured sequence of quantum gates applied to a set of qubits, analogous to a classical computational circuit. Quantum circuits provide a framework for implementation quantum algorithms and are typically represented using circuit diagrams, where qubits are depicted as horizontal lines and gates and gates as operations applied along these lines.

The execution of quantum circuit generally follows three stages:

1. **Initialization:** Qubits are presented in a known state, typically $|0\rangle$.
2. **Computation:** A sequence of quantum gates is applied to manipulate the state of the qubits.
3. **Measurement:** The quantum state is measured, collapsing it into a classical outcome.

Measurement plays a critical role in quantum computing, as it converts quantum information into classical data. However, measurement is inherently probabilistic and destroys the quantum state, which imposes constraints on how quantum algorithms are designed.

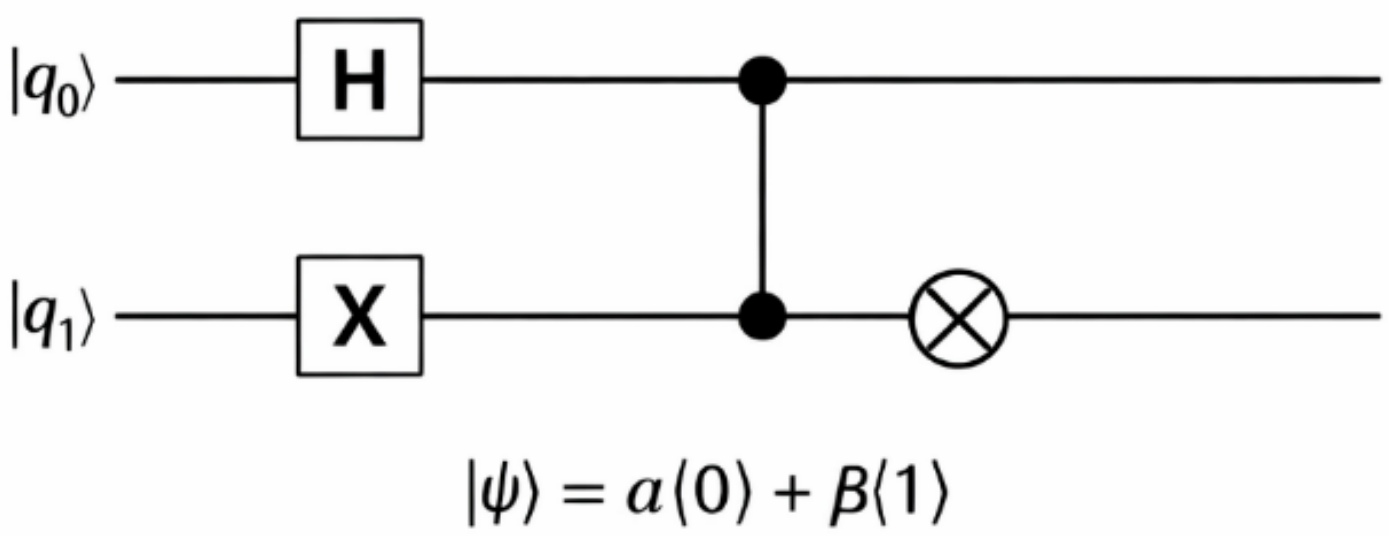

*Figure 7. A Basic Quantum Circuit with Common Gates*

Figure 7 illustrates a basic quantum circuit composed of commonly used quantum gates. It shows a simple two-qubit circuit. The first qubit $|q_0\rangle$ goes through a Hadamard gate, which creates a superposition, while the second qubit $|q_1\rangle$ passes through a Pauli-X gate, flipping its state. Afterwards, a CNOT gate entangles the two qubits, with $|q_0\rangle$ as the control and $|q_1\rangle$ as the target. The result is a combined quantum state $|\psi\rangle$ that represents a superposition and entanglement of the two qubits.

Quantum circuits can be broadly categorized into the following two types:

- *Deterministic circuits*, which aim to produce a specific output with high probability.
- *Variational circuits*, which include parameterized gates and are optimized using classical algorithms.

The latter category is particularly important for machine learning applications, as it enables the integration of quantum and classical optimization techniques (Benedetti eta al., 2019; Cerezo et al., 2021).

## Noisy Intermediate-Scale Quantum (NISQ) Devices

Current quantum hardware is characterized by the *noisy intermediate-scale quantum* (NISQ) regime, a term introduced by Preskill (Preskill, 2018). NISQ devices consist of tens of hundreds of qubits but are subject to significant noise, decoherence, and operational errors.

Some key challenges associated with NISQ devices include the following:

- *Decoherence*, where quantum states lose coherence due to interactions with the environment.
- *Gate errors*, arising from imperfect implementations of quantum operations.

Despite these limitations, NISQ devices have enabled the development of practical quantum algorithms that can be executed on near-term hardware. Hybrid quantum -classical approaches, which combine quantum circuits with classical optimization, have emerged as a promising strategy for leveraging NISQ systems (Bharti et al., 2022).

## Quantum Algorithms for Machine Learning

Quantum computing offers a variety of algorithms that have potential applications in various fields. These algorithms exploit quantum parallelism, interference, and entanglement to achieve computational advantages in specific tasks.

*Quantum Linear Algebra Algorithms:* Many machine learning algorithms rely on linear algebra operations such as matrix inversion, eigenvalue decomposition, and vector inner products. Quantum algorithms such as the *Harrow-Hassidim-Llyod* (HHL) algorithm provide exponential speedups for solving certain linear systems under specific conditions (Harrow et al., 2009; Biamonte et al., 2017). *Quantum Principal Component Analysis* (QPCA) enables the extraction of dominant eigenvectors from quantum-encoded data, facilitating dimensionality reduction in high-dimensional datasets (Lloyd et al., 2014).

*Quantum Search Algorithms*: Grover's algorithm is one of the most well-known quantum algorithms that offers a quadratic speedup for unstructured search problems (Grover, 1996; Biamonte et al., 2017). This capability can be applied to optimization and pattern matching tasks in machine learning, where large search space are common.

*Quantum Optimization Algorithms*: Optimization is central to machine learning, particularly in training models and tuning parameters. Quantum algorithms such as the quantum approximate optimization algorithm (QAOA) (Farhi et al., 2022) and quantum annealing methods (Neukart et al., 2017; Pomeroy et al, 2025; Date et al., 2019; Kim et al., 2025; Heidari et al., 2024; Yulianti et al., 2023l; Salloum et al., 2024; Kotsuki et al., 2024) haven been proposed for solving combinatorial optimization problems. These methods are particularly relevant for tasks such as feature selection, clustering, and hyperparameter optimization, where classical approaches may struggle with computational complexity.

*Variational Quantum Algorithms (VQAs):* Variational quantum algorithms are among the most promising approaches for near-term quantum machine learning. These algorithms use parameterized quantum circuits whose parameters are optimized using classical optimization techniques (Benedetti et al., 2017; Cerezo et al., 2021).

Examples of VQAs include the *variational quantum eigensolver* (VQE), the *quantum approximate optimization algorithm* (QAOA), and emerging models such as *quantum neural networks* (QNN).The VQE is primarily used for solving eigenvalue problems, particularly in quantum chemistry and materials science, where it estimates the ground-state energy of quantum systems through parameterized quantum circuits. The QAOA, on the other hand, is designed to address combinatorial optimization problems by iteratively improving candidate solutions using a hybrid quantum-classical optimization loop (Farhi et al., 2022). Quantum neural networks extend this paradigm by leveraging parameterized quantum circuits as trainable models for supervised and unsupervised learning tasks, enabling the representation of complex data distributions in quantum-enhanced feature spaces (Benedetti et al., 2017; Abbas et al., 2021).

VQAs are particularly well-suited for NISQ devices, as they rely on relatively shallow quantum circuits and incorporate classical optimization techniques to iteratively adjust circuit parameters. This hybrid approach allows them to tolerate noise and hardware imperfections that are characteristic of current quantum systems (Preskill, 2018; Bharti et al., 2022). Furthermore, VQAs form the foundation for hybrid quantum-classical learning frameworks, where quantum circuits are integrated with classical machine learning models to enhance computational efficiency and representational power (Abbas et al., 2021). Such architectures are increasingly being explored as practical pathways toward achieving quantum advantage in real-world machine learning applications.

## Quantum Data Encoding

One of the fundamental challenges in quantum machine learning is the efficient encoding of classical data into quantum states, a process often referred to as *quantum feature mapping or data embedding*. Since most real-world data is classical in nature, the effectiveness of any quantum algorithm critically depends on how this information is represented within a quantum system. Unlike classical computing, where data can be directly stored and manipulated, quantum systems require data to be encoded into quantum states before any computation can be performed. This encoding process not only influences the computational efficiency of quantum algorithms but also determines the expressive power and learning capability of quantum models (Biamonte et al., 2017; Schuld & Petruccione, 2018; Havlíček et al., 2019).

Several data encoding strategies have been proposed in the literature, each with its own advantages, limitations, and suitability for different types of applications. One of the simplest approaches is *basis encoding*, where classical binary data is directly mapped to the computational basis states of qubits, with each classical bit represented by a corresponding qubit state (Biamonte et al., 2017; Schuld & Petruccione, 2018; Benedetti et al., 2017). In this scheme, each classical bit is represented by a corresponding qubit state, such that a binary string can be encoded as a tensor product of individual qubit states. While basis encoding is straightforward and easy to implement, it is often inefficient in terms of qubit resources, as it requires one qubit per data dimension and does not exploit the full representational capacity of quantum systems (Schuld & Petruccione, 2018; Benedetti et al., 2017).

A more compact and theoretically powerful approach is *amplitude encoding*, in which classical data is encoded into the amplitudes of a quantum state vector. In this method, a normalized classical data vector is represented as a superposition of quantum basis states, enabling the encoding of $2^n$ data values using only $n$ qubits. This exponential compression makes amplitude encoding particularly attractive for high-dimensional data processing and has been widely studied in the context of quantum algorithms for linear algebra and machine learning, including quantum principal component analysis and quantum support vector machines (Lloyd et al., 2014; Rebentrost et al., 2014). However, the practical implementation of amplitude encoding can be challenging, as preparing arbitrary quantum states often requires complex quantum circuits and may incur significant overhead (Biamonte et al., 2017; Schuld & Petruccione, 2018).

Another widely used approach is *angle encoding* (also known as *parameter encoding*), where classical data values are encoded as rotation angles of quantum gates. In this scheme, input features are mapped to the parameters of single-qubit rotation gates (such as $R_x, R_y, R_z$), thereby embedding data into the quantum state through controlled transformations. Angle encoding is particularly suitable for variational quantum circuits and hybrid quantum-classical models, as it allows seamless integration with parameterized quantum circuits used in training processes (Bendetti et al., 2019; Cerezo et al., 2021). Compared to amplitude encoding, angle encoding is generally easier to implement on current quantum hardware and is more robust to noise, making it a practical choice for NISQ-era applications (Preskill, 2018; Bharti et al., 2022).

In addition to these basic encoding schemes, more advanced data embedding techniques have been proposed, including quantum feature maps in *quantum kernel methods* (Havlíček et al., 2019). These approaches transform classical data into high-dimensional quantum Hilbert spaces, where complex patterns may become more separable than in classical feature spaces. Such quantum-enhanced feature representations have been shown to improve the performance of certain machine learning tasks and form the basis of quantum kernel classifiers (Havlíček et al., 2019; Alexeev et al., 2025; Blank et al., 2020). The design of effective feature maps remains an active area of research, particularly in understanding how quantum advantage can be achieved through data encoding.

Efficient data encoding is therefore a critical factor in realizing the potential advantages of quantum computing. The choice of encoding strategy directly impacts the trade-offs between computational efficiency, circuit complexity, and learning performance. Poor encoding can negate any potential quantum advantage by introducing excessive overhead or limiting the expressiveness of the quantum model. Conversely, well-designed encoding schemes can enable quantum systems to process complex, high-dimensional data more effectively than classical counterparts. As a result, ongoing research continues to explore novel encoding techniques, optimization strategies, and hardware-aware implementations to bridge the gap between theoretical potential and practical deployment of quantum machine learning systems (Biamonte et al., 2017; Devadas & Sowmya, 2025; Lu et al., 2024; Melnikov et al., 2023).

***Summary:*** This section has introduced the fundamental concepts of quantum computing, including qubits, quantum gates, circuit algorithms, and data encoding strategies. These concepts form the technological foundation upon which quantum-enhanced artificial intelligence systems are built. The next section builds upon these principles to explore how quantum computing can be integrated with machine learning to develop advanced intelligent systems.

Key quantum computing concepts relevant to AI are summarized in Table 4.

*Table 4. fundamental Concepts in Quantum Computing*

| Concept | Description | Relevance to AI |
|---|---|---|
| Qubit | Quantum bit in superposition | Parallel computation |
| Entanglement | Correlated quantum states | Complex feature representation |
| Superposition | Multiple states simultaneously | High-dimensional search |
| Quantum Gates | Operations on qubits | Model transformations |
| Quantum Circuits | Sequence of gates | Learning architectures |

## FOUNDATIONS OF QUANTUM ARTIFICIAL INTELLIGENCE

Quantum Artificial Intelligence (QAI) represents an emerging interdisciplinary paradigm that integrates quantum computational techniques with artificial intelligence and data-driven modelling. By leveraging quantum mechanical principles such as superposition and entanglement, QAI aims to enhance learning efficiency, scalability, and robustness, particularly for complex and high-dimensional problem domains (Biamonte et al., 2017; Dunjko & Briegel, 2018; Devadas & Sowmya, 2025; Lu et al., 2024; Melnikov et al., 2023; Alexeev et al., 2025). While classical AI has achieved remarkable success across domains, its performance is fundamentally constrained by computational complexity, scalability issues, and limitations in optimization. Quantum computing, with its ability to process information in high-dimensional Hilbert spaces, offers a fundamentally different paradigm that may help overcome some of these challenges.

Quantum Machine Learning (QML), a core component of QAI, has evolved rapidly in recent years. Early foundational works in quantum machine learning established the theoretical basis for integrating quantum algorithms with learning frameworks (Biamonte et al., 2017; Dunjko & Briegel, 2018; Devadas & Sowmya, 2025; Lu et al., 2024; Melnikov et al., 2023). More recent studies have expanded this field significantly, exploring practical implementations of quantum-enhanced learning models, hybrid architectures, and domain-specific applications (Devadas & Sowmya, 2025; Lu et al., 2024; Melnikov et al., 2023; Alexeev et al., 2025). These developments suggest that QAI is not merely a theoretical construct but an emerging discipline with tangible implications for next-generation intelligent systems.

At its core, QAI seeks to answer a fundamental question: *how can quantum computational advantages be harnessed to improve the efficiency, scalability, and robustness of AI models*? Addressing this question requires a deep understanding of both quantum computing principles and machine learning methodologies.

## Quantum Machine Learning: Conceptual Foundations

*Quantum Machine Learning* (QML) focuses on the design of algorithms that utilize quantum resources to perform learning tasks. QML models aim to exploit properties such as superposition and entanglement to represent and process information in ways that are not feasible with classical systems.

Several quantum analogues of classical learning methods have been proposed in the literature. For instance, quantum versions of support vector machines (Rebentrost et al., 2014), principal component analysis (Llyod et al., 2014), and nearest neighbor methods (Wiebe et al., 2014) demonstrate how quantum algorithms can potentially accelerate core machine learning operations. These approaches often rely on encoding classical data into quantum states and performing transformations in exponentially large feature spaces.

A key theoretical insight in QML is that quantum systems can represent complex probability distributions more efficiently than classical systems in certain scenarios (Biamonte et al., 2017; Arunachalam & de Wolf, 2017). This capability is particularly relevant for high-dimensional data analysis, where classical methods often struggle with the curse of dimensionality. Additionally, quantum algorithms for linear algebra and optimization provide building blocks for scalable learning frameworks (Liu et al., 2024).

Some recent reviews (Devadas & Sowmya, 2025; Lu et al., 2024; Melnikov et al., 2023) emphasize that while many QML algorithms show theoretical advantages, their practical performance depends heavily on data encoding efficiency, circuit design, and hardware constraints (Devadas & Sowmya, 2025; Lu et al., 2024; Melnikov et al., 2023). Nonetheless, QML provides a promising foundation for exploring advanced AI capabilities.

## Variational Quantum Circuits and Parameterized Models

Variational Quantum Circuits (VQCs), also known as parameterized quantum circuits, are among the most widely studied approaches in QAI. These models consist of quantum circuits with tunable parameters that are optimized using classical algorithms. The hybrid nature of VQCs make them particularly suitable for *noisy intermediate-scale quantum* (NISQ) devices (Bharti et al., 2022).

In a typical VQC framework, a parameterized quantum circuit is used to transform input data encoded into quantum states. Measurement outcomes from the circuit are then used to compute a cost function, which is minimized using classical optimization techniques. This iterative process enables the model to learn task-specific representations.

Foundational work by Benedetti et al. (Bendetti et al., 2019) and comprehensive reviews by Cerezo et al. (Cerezo et al., 2021) have established VQCs as a flexible and powerful modeling paradigm. These circuits can be adapted for classification, regression, generative modeling, and reinforcement learning tasks. Furthermore, quantum neural networks, a subclass of VQCs, extend this framework by incorporating structures analogous to classical neural architectures (Abbas et al., 2021).

One of the strengths of VQCs lies in their expressivity. By exploiting entanglement and quantum interference, these circuits can represent complex functions that may be difficult to approximate classically. However, challenges such as barren plateaus (vanishing gradients) and optimization instability remain active areas of research.

## Quantum Kernel Methods and Feature Spaces

Quantum kernel methods provide another important approach within QAI. These methods leverage quantum feature maps to project classical data into high-dimensional quantum Hilbert spaces, where linear separation may become easier. The resulting kernels can then be used with classical algorithms such as support vector machines.

The seminal work by Havlíček et al. (Havlíček et al., 2019) demonstrated how quantum-enhanced feature spaces can improve classification performance for certain datasets. Subsequent research has further refined these methods, exploring tailored quantum kernels and efficient circuit implementations (Schuld & Petruccione, 2021; Blank et al., 2020).

The key advantage of quantum kernel methods lies in their ability to implicitly compute inner products in exponentially large feature spaces without explicitly constructing them. This property can lead to improved model expressiveness and better generalization in complex tasks.

However, the effectiveness of quantum kernels depends on the choice of feature map and the ability to efficiently estimate kernel values. Recent studies (Devadas & Sowmya, 2025; Lu et al., 2024) highlight that while quantum kernels show promise, their practical advantages over classical kernels remain context dependent.

## Quantum Optimization in Machine Learning

Optimization plays a central role in machine learning, and quantum computing offers several approaches for enhancing optimization processes. Quantum optimization algorithms aim to explore large solution spaces more efficiently than classical methods.

The Quantum Approximate Optimization Algorithm (QAOA) (Farhi et al., 2022) is one of the most prominent techniques in this domain. It is designed to solve combinatorial optimization problems by iteratively applying parameterized quantum operations. QAOA has been applied to problems such as graph optimization, scheduling, and feature selection.

Quantum annealing methods (Neukart et al., 2017; Pomeroy et al, 2025; Date et al., 2019; Kim et al., 2025; Heidari et al., 2024; Yulianti et al., 2023l; Salloum et al., 2024; Kotsuki et al., 2024) provide an alternative approach, leveraging adiabatic quantum evolution to find low-energy solutions of optimization problems formulated as Ising models (Lucas, 2014). These methods have been applied to clustering, classification, and ensemble learning tasks.

Recent advances (Liu et al., 2024; Iovane, 2025) suggest that quantum optimization may play a crucial role in training machine learning models, particularly in high-dimensional and non-convex landscapes. While current implementations are limited by hardware constraints, ongoing research continues to improve scalability and performance.

## Hybrid Quantum-Classical Learning Architectures

Given the limitations of current quantum hardware, hybrid quantum–classical architectures have emerged as a practical approach for implementing QAI systems. These architectures combine classical processing with quantum subroutines, allowing each component to handle tasks for which it is best suited.

In a typical hybrid model, classical neural networks are used for data preprocessing and feature extraction, while quantum circuits perform transformations or probabilistic modeling tasks. The outputs of quantum circuits are then integrated into classical learning pipelines.

A hybrid quantum-classical learning framework is illustrated in Figure 8. The process begins with classical data, such as an image of a cat, which cannot be directly processed by a quantum computer. This data is first encoded into quantum states using a quantum encoder, where classical information is mapped onto qubits. Once encoded, the data is passed through a variational quantum circuit, which contains parameterized quantum gates that transform the quantum state. These parameters are not fixed; they are tuned during the training to capture meaningful patterns. The circuit's output is then measured and fed into a classical optimizer, which evaluates the results and updates the parameters to improve the performance. This hybrid loop of quantum processing and classical optimization continues until the model reaches a good solution. Finally, the trained system produces an output, such as prediction or classification, which is expressed in classical form like a probability distribution or a bar chart. At the core of this workflow is the quantum state, represented as $|\psi\rangle = \alpha|0\rangle + \beta|1\rangle$, showing how information is stored in superposition and processed by the quantum circuit.

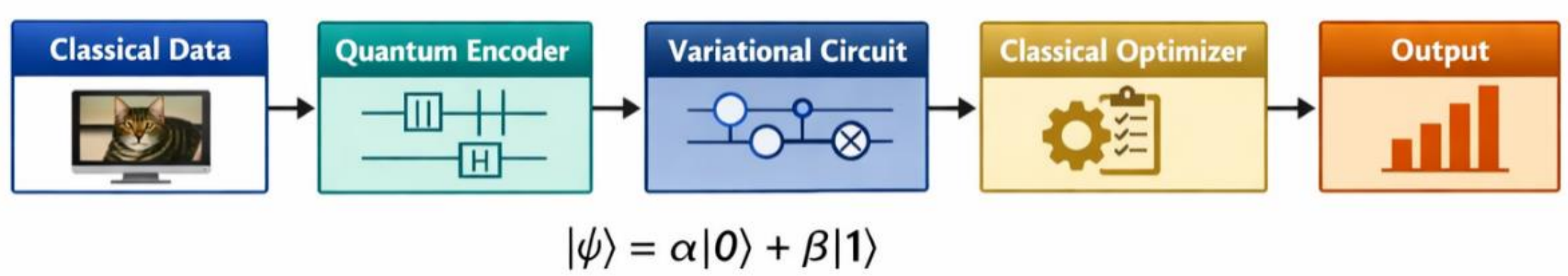

*Figure 8. Hybrid Quantum-Classical Learning Architecture*

Hybrid architectures are particularly effective in the NISQ era, where fully quantum systems are not yet feasible (Bharti et al., 2022). They also provide flexibility in model design, enabling researchers to experiment with different combinations of classical and quantum components.

Recent work (Huang et al., 2023) demonstrates the use of hybrid quantum-classical convolutional neural networks for image classification and robustness analysis. Similarly, studies (West et al., 2023a; West et al., 2023b; Gong et al., 2024; Dowling et al., 2026) have explored hybrid models for improving adversarial robustness, highlighting their potential in security-sensitive applications.

## Quantum Generative Models and Probabilistic Learning

Generative modeling is another area where quantum computing shows significant promise. Quantum generative models aim to learn probability distributions and generate new data samples using quantum circuits. Examples include quantum Boltzmann machines, quantum Born machines (Coyle et al., 2020), and variational quantum generative adversarial networks. These models leverage the probabilistic nature of quantum measurements to represent complex distributions efficiently.

Quantum generative models are particularly relevant for applications such as data synthesis, anomaly detection, and pattern recognition. Their ability to capture intricate correlations may provide advantages in scenarios where classical generative models struggle.

Recent studies (Melnikov et al., 2023; Orka et al., 2025) highlight the potential of quantum deep learning approaches in domains such as neuroinformatics and healthcare analytics. However, practical implementation challenges, including data encoding and noise sensitivity, remain key obstacles.

Table 5 provides an overview of key quantum machine learning models and their applications.

*Table 5. Quantum Machine Learning Models and Techniques*

| Model Type | Description | Example References | Application |
|---|---|---|---|
| Variation Quantum Circuits | Parameterized circuits | Cerezo et al. (2021) | Optimization |
| Quantum Kernel Methods | Feature mapping in Hillbert space | Havlíček et al. (2019) | Classification |
| Quantum Neural Networks | Quantum analog of neural nets | Abbas et al. (2021) | Pattern recognition |
| Quantum Annealing | Optimization via energy minimization | Neukart et al. (2017) | Combinatorial problems |

## Relevance to Adversarial Robustness

The principles of QAI have important implications for adversarial machine learning. Quantum feature spaces, probabilistic modeling, and optimization techniques may provide new ways to improve model robustness and detect adversarial perturbations.

For instance, quantum-enhanced feature representations may increase class separability, reducing vulnerability to adversarial manipulation. Similarly, quantum optimization algorithms may enable more effective adversarial training by exploring diverse perturbation scenarios.

Recent studies (West et al., 2023a; West et al., 2023b; Gong et al., 2024; Dowling et al., 2026; Huang et al., 2023; Nowmi et al.) provide early evidence that quantum machine learning models may exhibit improved robustness under certain conditions. Hybrid architectures further enhance this potential by combining quantum transformations with classical defenses.

Although these approaches are still in early stages, they represent a promising direction for developing secure and trustworthy AI systems. While significant challenges remain, the integration of quantum computing with artificial intelligence offers a significant pathway for addressing limitations in classical AI systems. In particular, the potential for enhanced robustness, scalability, and learning efficiency makes QAI a critical area of research for next-generation intelligent systems.

# QUANTUM-ENHANCED ADVERSARIAL ROBUSTNESS IN AI

The increasing vulnerability of modern machine learning systems to adversarial attacks has raised serious concerns regarding the reliability and security of artificial intelligence in real-world applications. While classical defense mechanisms such as adversarial training, input preprocessing, and robust optimization have demonstrated partial success, they often struggle to provide generalizable protection against adaptive and sophisticated adversarial strategies (Madry et al., 2018; Biggio & Roli, 2018; Chakraborty et al., 2021; Han et al., 2023). These limitations highlight the need for fundamentally new computational paradigms capable of addressing the inherent weaknesses of classical learning systems. In this context, quantum computing presents a promising direction for enhancing adversarial robustness by leveraging its unique computational properties, including superposition, entanglement, and probabilistic state evolution (Nielsen & Chuang, 2010; Preskill, 2018; Bharti et al., 2022).

Quantum Artificial Intelligence (QAI) offers the potential to redefine how machine learning models are trained, optimized, and secured. By integrating quantum computational techniques with classical learning frameworks, it becomes possible to explore high-dimensional feature spaces more effectively, perform optimization tasks more efficiently, and develop probabilistic models that capture complex data distributions with greater fidelity (Biamonte et al., 2017; Dunjko & Briegel, 2018; Devadas & Sowmya, 2025; Lu et al., 2024; Melnikov et al., 2023; Alexeev et al., 2025). These capabilities are particularly relevant for adversarial machine learning, where robustness depends on the ability to generalize beyond training distributions and resist carefully crafted perturbations.

This section presents a comprehensive exploration of conceptual frameworks through which quantum computing can enhance adversarial robustness in AI systems. The discussion focuses on three primary dimensions: (i) quantum optimization for robust learning, (ii) quantum-enhanced feature representations, and (iii) hybrid quantum–classical architectures for secure AI systems. Additionally, the section examines quantum-based adversarial detection mechanisms and outlines emerging research directions that may shape the future of secure AI.

## Quantum Optimization for Robust Learning

Optimization lies at the core of machine learning, governing processes such as parameter estimation, loss minimization, and adversarial training. Many adversarial attacks exploit weaknesses in optimization landscapes, identifying directions in which small perturbations can cause significant degradation in model performance (Goodfellow et al., 2015; Madry et al., 2018; Biggio & Roli, 2018). Consequently, improving optimization strategies is critical for enhancing robustness.

Quantum optimization algorithms offer a fundamentally different approach to exploring solution spaces. Techniques such as the *quantum approximate optimization algorithm* (QAOA) and quantum annealing are designed to efficiently search complex, high-dimensional spaces for optimal or near-optimal solutions (Farhi et al., 2022; Neukart et al., 2017; Pomeroy et al., 2025; Date et al., 2019; Kim et al., 2025; Heidari et al., 2024; Yulianti et al., 2023; Salloum et al., 2024; Kotsuki et al., 2024). Unlike classical optimization methods, which often rely on gradient-based updates and may become trapped in local minima, quantum algorithms can exploit quantum superposition to evaluate multiple candidate solutions simultaneously.

QAOA, for instance, operates by preparing a parameterized quantum state that encodes the solution space of a combinatorial optimization problem and iteratively refining this state using classical optimization loops (Farhi et al., 2022; Bharti et al., 2022). This hybrid quantum–classical approach allows QAOA to approximate optimal solutions for problems that are computationally intractable for classical algorithms. In the context of adversarial machine learning, QAOA can be used to formulate adversarial training as a robust optimization problem, where the objective is to minimize worst-case loss under adversarial perturbations.

Similarly, quantum annealing provides a mechanism for solving optimization problems by encoding them into energy landscapes and allowing the system to evolve toward low-energy configurations (Neukart et al., 2017; Pomeroy et al., 2025; Date et al., 2019; Kim et al., 2025; Heidari et al., 2024; Yulianti et al., 2023; Salloum et al., 2024; Kotsuki et al., 2024). This approach has been applied to feature selection, clustering, and ensemble learning tasks, demonstrating its potential for improving model robustness (Pomeroy et al. 2025; Yulianti et al., 2023). By enabling more effective exploration of solution spaces, quantum annealing may help identify model parameters that are inherently more resistant to adversarial perturbations.

Recent studies have explored the role of quantum optimization in adversarial robustness. For example, research has shown that quantum-enhanced optimization techniques can improve the stability of learning processes and reduce susceptibility to adversarial manipulation (West et al., 2023a; West et al., 2023b; Dowling et al., 2026). These findings suggest that quantum optimization could play a critical role in developing next-generation robust learning frameworks.

## Quantum-Enhanced Feature Representations

Another fundamental factor contributing to adversarial vulnerability is the representation of data in feature space. Classical machine learning models often rely on linear or nonlinear transformations that may fail to capture complex relationships in high-dimensional data, making them susceptible to adversarial perturbations.

Quantum computing introduces new possibilities for feature representation using quantum states and feature maps. Quantum feature mapping involves encoding classical data into quantum states in a high-dimensional Hilbert space, where complex patterns can be more easily separated (Havlíček et al., 2019; Schuld & Petruccione, 2021; Blank et al., 2020). This approach is analogous to kernel methods in classical machine learning but operates in exponentially large feature spaces that are difficult to simulate classically.

Quantum kernel methods have demonstrated promising results in classification tasks by leveraging quantum-enhanced feature spaces (Havlíček et al., 2019; Schuld & Petruccione, 2021). These methods compute similarity measures between data points using quantum circuits, enabling the construction of decision boundaries that are more expressive and potentially more robust to adversarial perturbations. By mapping data into higher-dimensional spaces, quantum feature representations can reduce the effectiveness of small input perturbations, thereby improving robustness.

In addition to kernel methods, *variational quantum circuits* (VQCs) provide a flexible framework for learning feature representations directly from data (Benedetti et al., 2019; Cerezo et al., 2021). These circuits consist of parameterized quantum gates that can be trained using classical optimization techniques to perform specific tasks such as classification or regression. Because VQCs operate in quantum state spaces, they can capture complex correlations and dependencies that are difficult to model using classical architectures.

The use of quantum feature representations also aligns with recent research on adversarial robustness, which suggests that models with richer and more structured feature spaces are less susceptible to adversarial manipulation (Han et al., 2023; West et al., 2023b). By leveraging quantum-enhanced representations, it may be possible to develop models that generalize better and exhibit improved resistance to adversarial attacks.

## Hybrid Quantum-Classical Learning Architectures

Given the current limitations of quantum hardware, fully quantum machine learning systems remain impractical for large-scale applications. Instead, hybrid quantum–classical architectures have emerged as a promising approach for integrating quantum computing into AI systems (Bharti et al., 2022; Cerezo et al., 2021; Coyle et al., 2020).

In these architectures, quantum circuits are used as components within classical learning frameworks, enabling the combination of quantum computational advantages with the scalability of classical systems. For example, a hybrid model may use a classical neural network for feature extraction, followed by a quantum circuit for classification or decision-making. Alternatively, quantum circuits may be used to perform specific sub-tasks such as optimization, sampling, or probabilistic modeling.

Hybrid models are particularly well-suited for Noisy Intermediate-Scale Quantum (NISQ) devices, as they require relatively shallow quantum circuits and can tolerate noise through classical feedback loops (Preskill, 2018; Bharti et al., 2022). Variational quantum algorithms, including VQE and QAOA, play a central role in these architectures by enabling parameter optimization through iterative quantum–classical interactions.

Recent studies have demonstrated the potential of hybrid models for improving adversarial robustness. For instance, hybrid quantum–classical convolutional neural networks have been shown to exhibit enhanced robustness against adversarial attacks compared to purely classical models (Huang et al., 2023). Similarly, research on quantum neural networks suggests that the probabilistic nature of quantum systems may introduce inherent resistance to certain types of adversarial perturbations (Abbas et al, 2021).

The integration of quantum and classical components also enables the development of adaptive defense mechanisms. By dynamically adjusting quantum circuit parameters based on observed adversarial behavior, hybrid systems can potentially respond to attacks in real time, improving resilience in adversarial environments.

## Quantum-Based Adversarial Detection Mechanisms

In addition to improving model robustness, quantum computing offers new opportunities for detecting adversarial inputs. Adversarial detection involves identifying whether a given input has been manipulated before it is processed by the model.

Quantum probabilistic models and anomaly detection techniques provide a promising framework for this task. Because quantum systems naturally represent probability distributions, they can be used to model complex data distributions and identify deviations that may indicate adversarial manipulation (Biamonte et al., 2017; Benedetti et al., 2019; Arunachalam & de Wolf, 2017; Devadas & Sowmya, 2025). By analyzing quantum state distributions, it may be possible to detect subtle anomalies that are difficult to capture using classical methods.

Quantum kernel methods can also be applied to anomaly detection by measuring distances between data points in quantum feature spaces. Inputs that deviate significantly from the training distribution can be flagged as potential adversarial examples. This approach leverages the high-dimensional nature of quantum feature spaces to improve detection accuracy.

Recent research has explored the use of quantum techniques for adversarial detection. For example, randomized quantum encoding methods have been proposed to enhance robustness and detect adversarial inputs by introducing stochasticity into feature representations (Gong et al., 2024). Similarly, benchmarking studies have demonstrated that quantum models can achieve competitive performance in adversarial detection tasks (West et al., 2023a).

Although these approaches are still in early stages of development, they highlight the potential of quantum computing to provide new layers of security for AI systems.

## Conceptual Framework for Quantum-Enhanced Adversarial Robustness

Building on the preceding discussions, a unified conceptual framework for quantum-enhanced adversarial robustness can be proposed. This framework integrates quantum optimization, feature representation, and hybrid learning architectures into a cohesive approach for developing secure AI systems.

At the core of this framework is the idea of combining quantum-enhanced learning processes with classical defense strategies. Quantum optimization algorithms can be used to improve adversarial training by identifying robust model parameters. Quantum feature mapping techniques can enhance data representations, making it more difficult for adversarial perturbations to affect model predictions. Hybrid architectures can integrate these components into scalable systems capable of operating in real-world environments.

The framework also incorporates quantum-based detection mechanisms to provide additional layers of security. By combining robustness and detection, AI systems can achieve a more comprehensive defense against adversarial threats.

Recent studies have begun to validate elements of this framework. Research on quantum adversarial robustness has demonstrated that quantum models can achieve improved resistance to adversarial attacks under certain conditions (West et al., 2023a; West et al., 2023b; Gong et al., 2024; Dowling et al., 2026; Huang et al., 2023; Nowmi et al., 2025). However, further work is needed to fully understand the theoretical foundations and practical implications of these approaches.

Figure 9 presents the conceptual framework for quantum-enhanced adversarial robustness. The process begins with classical input data (illustrated with the cat image), which is then transformed into quantum states through a *quantum encoder* using gates like Hadamard. The encoded quantum information is processed by a *quantum model*, where computations occur in the quantum circuit. To strengthen resilience against malicious perturbations, the system incorporates an *adversarial training loop* that introduces small disturbances (denoted as +δ) and iteratively updates the model to improve robustness. Finally, the output passes through a *detection module*, symbolized by a shield and magnifying glass, which ensures security and verifies the integrity of the results. The quantum state equation $|\psi\rangle = \alpha|0\rangle + \beta|1\rangle$, highlights the fundamental representation of quantum information underlying the entire pipeline.

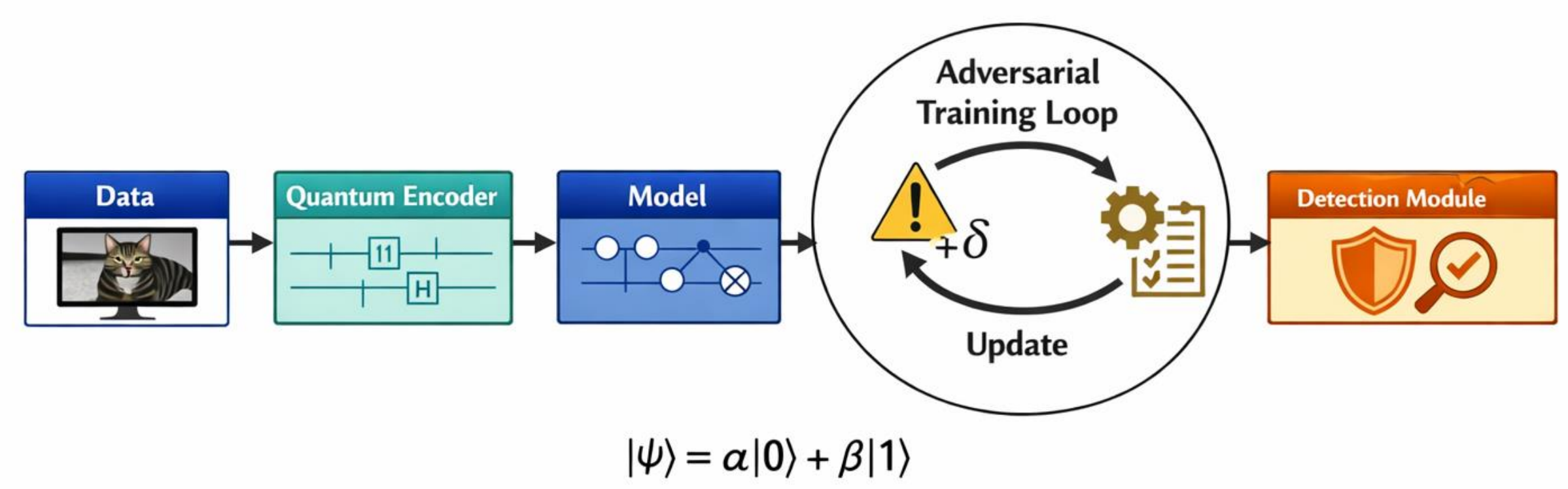

*Figure 9. Conceptual Framework for Quantum-Enhanced Adversarial Robustness*

Table 6 summarizes the key quantum-enhanced strategies for improving adversarial robustness.

*Table 6. Quantum-Enhanced Approaches for Adversarial Robustness*

| Approach | Mechanism | Advantage | Limitation |
|---|---|---|---|
| Quantum Optimization | Improved adversarial | Better minima | Hardware constraints |
| Quantum Feature Mapping | High-dimensional encoding | Improved separability | Encoding complexity |
| Hybrid Models | Classical + quantum layers | Practical implementation | Integration challenges |
| Quantum Detection | Anomaly detection | Beter detection | Early-stage research |

## Future Directions in Quantum Adversarial Robustness

Despite the promising potential of quantum-enhanced adversarial robustness, several challenges must be addressed before these techniques can be widely adopted. One of the primary limitations is the current state of quantum hardware, which is characterized by noise, limited qubit counts, and short coherence times (Preskill, 2018; Bharti et al., 2022). These constraints restrict the complexity of quantum algorithms that can be implemented in practice.

Another challenge lies in the development of efficient data encoding methods. Encoding classical data into quantum states remains a non-trivial problem, and inefficient encoding can negate potential quantum advantages (Schuld & Petruccione, 2018; Schuld & Petruccione, 2021; Havlíček et al., 2019). Additionally, the training of variational quantum circuits is often affected by issues such as barren plateaus, where gradients vanish and optimization becomes difficult (Cerezo et al., 2021).

From a theoretical perspective, the extent to which quantum computing can improve adversarial robustness is not yet fully understood. While initial studies are promising, more rigorous analysis is needed to establish formal guarantees and identify the conditions under which quantum advantages can be realized. Moreover, integrating quantum and classical systems presents practical challenges related to system design, scalability, and interoperability. Addressing these challenges will require collaboration across disciplines, including quantum computing, machine learning, and cybersecurity.

Future research directions include the development of scalable quantum optimization algorithms for adversarial training, the design of robust quantum feature maps, and the exploration of new hybrid architectures. Another promising direction is the study of quantum adversarial attacks, which could provide insights into the limitations of quantum models and inform the design of more robust defenses. Additionally, advances in quantum hardware and error correction may enable the implementation of more complex quantum machine learning models, further expanding the scope of QAI.

Interdisciplinary collaboration will play a critical role in advancing this field. By combining expertise from quantum physics, computer science, and cybersecurity, researchers can develop innovative solutions that address the challenges of adversarial robustness in AI systems.

## APPLICATIONS OF QUANTUM-ENHANCED ROBUST AI SYSTEMS

The integration of quantum computing with artificial intelligence is rapidly transforming the landscape of intelligent systems, particularly in domains where robustness, reliability, and security are critical. As adversarial machine learning continues to expose the vulnerabilities of classical AI systems, the emergence of quantum-enhanced techniques offers a promising pathway toward the development of resilient and trustworthy AI frameworks. Quantum-enhanced robust AI systems combine the computational advantages of quantum algorithms with advanced machine learning techniques to address challenges related to scalability, adversarial robustness, and secure decision-making (Devadas & Sowmya, 2025; Lu et al., 2024; Melnikov et al., 2023; Alexeev et al., 2025; Liu et al., 2024; Iovane, 2025).

Recent advances in quantum machine learning (QML) have demonstrated that quantum models can achieve improved robustness under adversarial settings, particularly when leveraging hybrid architectures and probabilistic representations (West et al., 2023a; West et al., 2023b; Gong et al., 2024; Dowling et al., 2026; Huang et al., 2023; Nowmi et al., 2025). Furthermore, emerging research indicates that noise characteristics in Noisy Intermediate-Scale Quantum (NISQ) devices may even act as a natural regularization mechanism, potentially enhancing resistance to adversarial perturbations under certain conditions (Nowmi et al., 2025).

This section presents a comprehensive discussion of practical and emerging applications of quantum-enhanced robust AI across multiple domains, including healthcare, finance, transportation, cybersecurity, and intelligent infrastructure, along with forward-looking perspectives on futuristic applications enabled by this convergence. The broad application landscape of quantum-enhanced robust AI is shown in Figure 10.

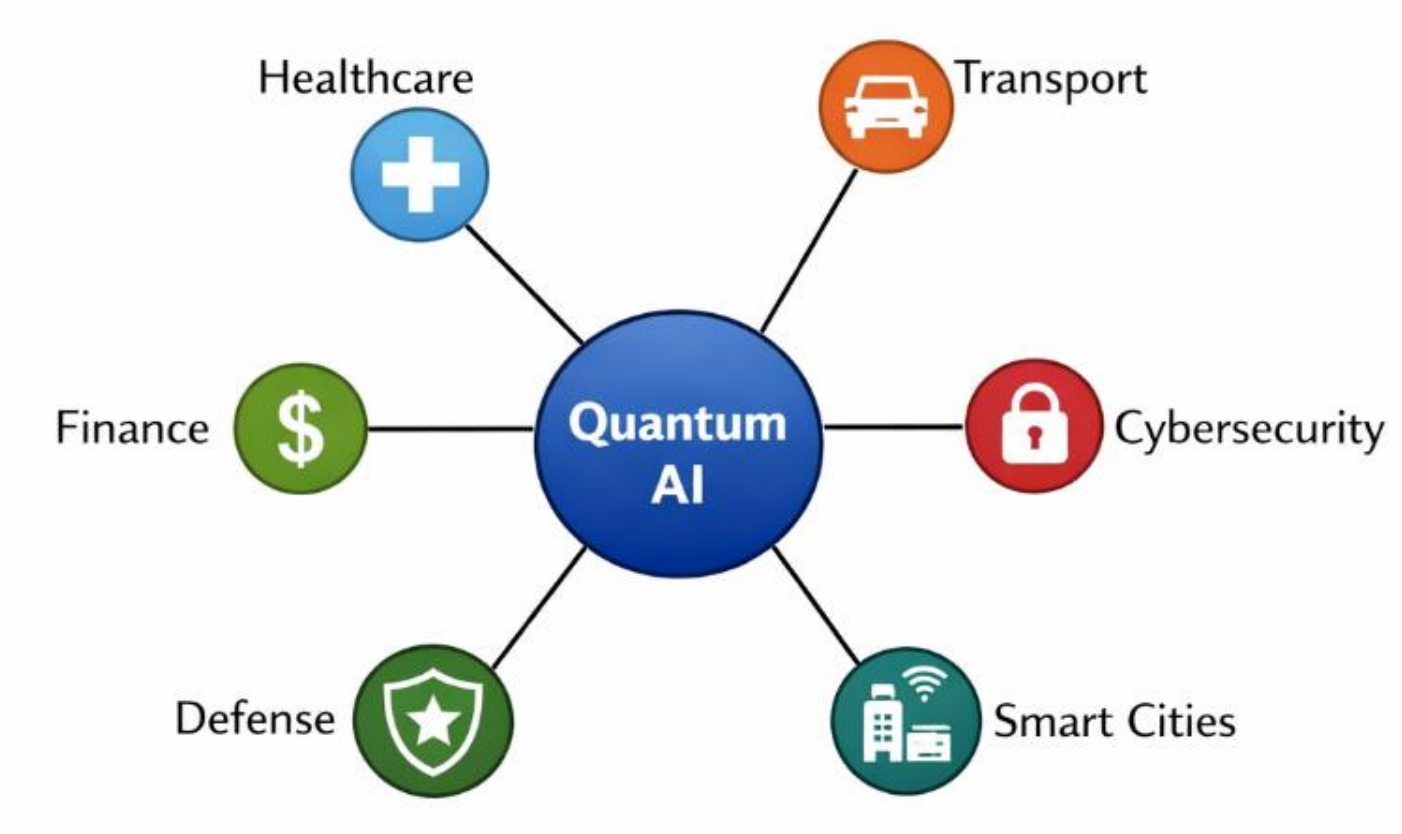

*Figure 10. Application Domains of Quantum-Enhanced Robust AI*

## Healthcare and Medical Diagnostics

Healthcare represents one of the most impactful domains for AI deployment, where model reliability and robustness are essential due to the high stakes involved in clinical decision-making. AI systems are increasingly used for medical imaging, diagnostics, predictive analytics, and personalized medicine. However, adversarial vulnerabilities in these systems pose serious risks, as even minor perturbations in input data can lead to incorrect diagnoses.

Quantum-enhanced robust AI offers significant advantages in this domain by enabling improved representation and processing of high-dimensional biomedical data. Quantum machine learning models can exploit Hilbert space representations to capture complex correlations in medical datasets, thereby improving classification accuracy and robustness (Havlíček et al., 2019; Devadas & Sowmya, 2025; Lu et al., 2024; Melnikov et al., 2023). Recent systematic studies have highlighted the growing role of QML in healthcare, particularly in medical imaging, diagnostic prediction, and clinical decision support (Gupta et al., 2025).

Quantum-enhanced feature mapping techniques can reduce the susceptibility of models to adversarial perturbations by increasing the separability of data in high-dimensional feature spaces. Additionally, quantum optimization algorithms can improve adversarial training processes in medical AI systems, enabling more robust model parameter selection. These approaches are particularly relevant in applications such as cancer detection, radiology, and genomics, where robustness is critical.

Furthermore, quantum computing is expected to play a transformative role in drug discovery and molecular modeling. The ability to simulate quantum interactions at the molecular level enables more accurate modeling of chemical processes, while robust AI ensures that predictions are not influenced by adversarial manipulation. Recent research also emphasizes the integration of quantum algorithms with healthcare data pipelines to enhance both computational efficiency and security (Marengo & Santamato, 2025).

In future, quantum-enhanced AI systems may enable fully autonomous clinical decision-support systems that operate securely in real-time environments, providing reliable diagnostics while ensuring data integrity and resilience against adversarial threats (Gupta et al., 2025; Marengo & Santamato, 2025; Catak et al., 2025; Hai et al., 2025; Innan et al., 2025; Dowling et al., 2026; Klymenko et al., 2025).

## Financial Systems and Fraud Detection

The financial sector is highly dependent on AI for fraud detection, credit scoring, risk assessment, and algorithmic trading. However, financial AI systems are particularly vulnerable to adversarial manipulation, which can lead to significant economic losses and systemic risks.

Quantum-enhanced robust AI introduces new capabilities for financial analytics by enabling more efficient optimization and improved robustness against adversarial inputs. Quantum optimization techniques, such as QAOA and quantum annealing, can be applied to portfolio optimization and risk management problems, allowing for more efficient exploration of complex financial landscapes (Farhi et al., 2022; Neukart et al., 2017; Pomeroy et al., 2025; Date et al., 2019; Kim et al., 2025; Heidari et al., 2024; Yulianti et al., 2023; Salloum et al., 2024; Kotsuki et al., 2024).

Recent studies demonstrate that quantum machine learning models, such as *quantum* LSTM (QLSTM) and *quantum support vector machines* (QSVM), can outperform classical models in financial forecasting while maintaining higher resilience to adversarial attacks (Tiwo, 2025). These models exhibit lower degradation in performance under adversarial conditions, highlighting their potential for secure financial analytics.

Additionally, quantum computing is being explored for applications in blockchain security and quantum-resistant cryptographic systems, which are essential for safeguarding financial transactions in the quantum era (Naik et al., 2025). The integration of quantum-enhanced AI with secure communication protocols such as quantum key distribution (QKD) can further strengthen financial cybersecurity frameworks.

In algorithmic trading, quantum-enhanced AI can enable faster analysis of market data and more robust trading strategies. By incorporating adversarial robustness, these systems can better withstand attempts to manipulate market signals or exploit vulnerabilities in predictive models.

Future financial ecosystems may leverage quantum-enhanced AI to enable real-time fraud detection, adaptive risk management, and secure *decentralized finance* (DeFi) platforms that are resilient to both classical and quantum threats (Yu & Luo, 2025; Cardaioli et al., 2025; Abbou et al., 2025; Weinberg & Faccia, 2024; Thakkar et al., 2024; Sawaika et al., 2025; Pahuja, 2025; Yu & Shen, 2025)

## Autonomous Transportation and Intelligent Mobility

Autonomous transportation systems rely on artificial intelligence for perception, decision-making, and control. These systems operate in highly dynamic and safety-critical environments, making them particularly vulnerable to adversarial attacks that can manipulate sensor inputs or disrupt decision processes (Biggio & Roli, 2018; Chakraborty et al., 2021; Han et al., 2023; Wendlinger et al., 2024). Recent studies have demonstrated that adversarial perturbations in images, LiDAR signals, or communication channels can significantly degrade the performance of autonomous driving systems, raising serious concerns regarding safety and reliability (Han et al., 2023; Waghela et al., 2024c; Waghela et al., 2024d; Waghela et al., 2024f).

Quantum-enhanced robust AI has the potential to significantly improve the reliability of autonomous systems by enhancing both data processing and decision-making capabilities. In particular, quantum machine learning models can enable more expressive feature representations, allowing systems to better distinguish between legitimate and adversarial inputs (Biamonte et al., 2017; Devadas & Sowmya, 2025; Lu et al., 2024; West et al., 2023b). Quantum feature mapping techniques, such as those used in quantum kernel methods, can transform sensor data into higher-dimensional Hilbert spaces where class separability is improved, thereby reducing susceptibility to adversarial perturbations (Havlíček et al., 2019; Schuld & Petruccione, 2021; Blank et al., 2020; West et al., 2023a).

In addition to perception, optimization plays a critical role in autonomous transportation, especially in route planning, traffic management, and resource allocation. Quantum optimization algorithms, including the Quantum Approximate Optimization Algorithm (QAOA) and quantum annealing methods, provide promising approaches for solving large-scale combinatorial optimization problems more efficiently than classical techniques (Farhi et al., 2022; Liu et al., 2024; Neukart et al., 2017; Pomeroy et al., 2025; Date et al., 2019; Kim et al., 2025; Heidari et al., 2024; Yulianti et al., 2023; Salloum et al., 2024; Kotsuki et al., 2024). These methods can be applied to intelligent traffic systems to optimize routing decisions in real time, reduce congestion, and improve energy efficiency in transportation networks (Kim et al., 2025; Heidari et al., 2024; Kotsuki et al., 2024).

Hybrid quantum–classical architectures further extend these capabilities by integrating classical perception modules with quantum-enhanced decision-making components. Such architectures are particularly suitable for near-term Noisy Intermediate-Scale Quantum (NISQ) devices, where quantum circuits can be used for specialized tasks such as probabilistic reasoning, optimization, or anomaly detection, while classical systems handle large-scale data processing (Benedetti et al., 2019; Cerezo et al., 2021; Abbas et al. 2021). Recent research has also shown that hybrid quantum-classical models can improve robustness in classification and decision-making tasks, including applications in image-based systems relevant to autonomous driving (Huang et al., 2023; West et al., 2023a).

Another important dimension is the resilience of autonomous systems against adversarial manipulation. Quantum-enhanced learning frameworks have been explored for improving adversarial robustness by enabling more effective training strategies and detection mechanisms (West et al., 2023a; West et al., 2023b; Gong et al. 2024; Dowling et al., 2024; Huang et al. 2023; Nowmi et al., 2025). These approaches may allow autonomous systems to identify anomalous patterns in sensor data and respond to adversarial inputs more effectively, thereby enhancing operational safety.

Future transportation systems may incorporate quantum-enhanced AI to create fully autonomous, secure, and self-optimizing mobility networks. These systems could leverage advances in quantum communication and quantum-safe cryptographic techniques to ensure secure data exchange between vehicles, infrastructure, and control systems, thereby reducing the risk of cyberattacks and data manipulation (Preskill, 2018; Alexeev et al., 2025; Sen, 2025). Furthermore, the integration of quantum computing with edge AI systems may enable real-time adaptive decision-making in complex urban environments, supporting the development of intelligent transportation ecosystems that are not only efficient but also resilient and trustworthy.

## Cybersecurity and Threat Intelligence

Cybersecurity represents one of the most natural and critical application domains for quantum-enhanced robust AI, as it inherently operates in adversarial environments where attackers continuously attempt to exploit system vulnerabilities. Artificial intelligence is widely deployed in cybersecurity tasks such as intrusion detection, malware classification, anomaly detection, and threat intelligence analysis. However, these AI-driven systems themselves are susceptible to adversarial manipulation, where carefully crafted inputs can evade detection mechanisms or trigger incorrect classifications (Biggio & Roli, 2018; Chakraborty et al., 2021; Han et al., 2023). Such vulnerabilities pose serious risks, particularly in mission-critical infrastructures where the integrity and reliability of security systems are paramount.

Quantum-enhanced AI offers a promising pathway for strengthening cybersecurity systems by enabling more powerful data analysis and robust learning mechanisms. Quantum machine learning models are capable of operating in high-dimensional Hilbert spaces, allowing them to capture intricate correlations and patterns in network traffic, system logs, and behavioral data that may be difficult to identify using classical techniques (Biamonte et al., 2017; Devadas & Sowmya, 2025; Lu et al., 2024; Melnikov et al., 2023). For example, quantum kernel methods and variational quantum circuits can provide richer feature representations, improving the detection of subtle anomalies and previously unseen attack patterns (Havlíček et al., 2019; Schuld & Petruccione, 2021; Blank et al., 2020; Cerezo et al., 2021). This capability is particularly valuable in modern cybersecurity scenarios, where attacks are increasingly sophisticated, polymorphic, and difficult to detect using signature-based approaches.

In addition to enhanced detection capabilities, quantum computing also plays a significant role in securing the underlying infrastructure of AI systems. Recent research has proposed multi-layered frameworks for post-quantum secure machine learning that integrate quantum-resistant cryptographic techniques with AI pipelines to ensure end-to-end security (Alexeev et al., 2025; Sen, 2025). These frameworks combine advanced technologies such as post-quantum cryptography, fully homomorphic encryption, and distributed ledger systems (e.g., blockchain) to protect sensitive data, model parameters, and communication channels from both classical and quantum-enabled adversaries. Such integrated approaches are essential in future cybersecurity ecosystems, where quantum computing may simultaneously act as both a tool for defense and a potential threat to existing cryptographic protocols.

Another important dimension of quantum-enhanced cybersecurity lies in the robustness of machine learning models themselves. Recent studies in quantum adversarial machine learning have demonstrated that quantum classifiers may exhibit inherent resilience to certain types of adversarial attacks due to properties such as probabilistic state representation, superposition, and non-linear feature mappings (West et al., 2023a; West et al., 2023b; Dowling et al., 2024). These characteristics can make it more difficult for adversaries to construct effective perturbations, particularly in high-dimensional quantum feature spaces. Furthermore, theoretical investigations into adversarial robustness in quantum machine learning have provided formal bounds and guarantees on model behavior under adversarial conditions, offering new insights into the design of secure learning systems (Gong et al., 2024; Dowling et al., 2024; Nowmi et al., 2025).

Hybrid quantum–classical architectures further enhance cybersecurity capabilities by combining the strengths of classical data processing with quantum-enhanced analysis. In such systems, classical components handle large-scale data ingestion and preprocessing, while quantum modules perform specialized tasks such as anomaly detection, probabilistic inference, or optimization (Benedetti et al., 2019; Cerezo et al., 2021; Abbas et al., 2021). This hybrid approach is particularly suitable for deployment in real-world environments, where current quantum hardware limitations necessitate efficient integration with classical systems. Moreover, recent work has shown that hybrid models can improve adversarial robustness and detection accuracy in security-critical applications (West et al., 2023a; Huang et al., 2023).

Future cybersecurity systems are expected to evolve toward fully autonomous, intelligent, and self-adaptive defense frameworks powered by quantum-enhanced AI. These systems may enable real-time threat detection and response, continuous monitoring of complex networks, and predictive threat intelligence capable of anticipating attacks before they occur. Additionally, the integration of quantum communication technologies, such as quantum key distribution (QKD), can provide provably secure communication channels that are resistant to eavesdropping and cryptographic attacks (Nielsen & Chuang, 2010; Preskill, 2018; Sen, 2025).

Furthermore, the convergence of quantum computing, AI, and cybersecurity may lead to the development of self-healing systems that can automatically detect vulnerabilities, adapt to evolving threats, and reconfigure themselves to maintain security and performance. Such systems will be particularly important in critical infrastructures, including smart grids, defense systems, financial networks, and cloud computing platforms, where resilience against adversarial attacks is essential.

In summary, quantum-enhanced robust AI has the potential to redefine the landscape of cybersecurity and threat intelligence by enabling more powerful detection mechanisms, stronger cryptographic protections, and inherently robust learning models. While significant challenges remain in terms of scalability, hardware limitations, and practical deployment, ongoing research at the intersection of quantum computing and AI security is likely to play a pivotal role in shaping next-generation cybersecurity solutions.

## Smart Infrastructure and Urban Systems

Smart cities and intelligent infrastructure systems represent a transformative application of artificial intelligence, enabling the efficient management of urban resources, optimization of public services, and enhancement of quality of life. AI-driven systems are widely used in domains such as energy management, transportation, water distribution, waste management, and public safety. These systems rely on continuous data collection and real-time decision-making, often operating across interconnected networks of sensors, devices, and control systems. However, the increasing reliance on AI and interconnected digital infrastructure also introduces significant vulnerabilities, particularly in the form of cyber threats and adversarial attacks (Biggio & Roli, 2018; Chakraborty et al., 2021; Han et al., 2023). As a result, ensuring robustness, security, and resilience has become a critical requirement for the sustainable development of smart infrastructure systems.

Quantum-enhanced robust AI offers a promising avenue for addressing these challenges by enabling more efficient computation, improved optimization, and stronger resilience against adversarial manipulation. In particular, quantum machine learning techniques can enhance the analysis of complex, high-dimensional urban data, allowing systems to identify patterns and anomalies that may not be easily detectable using classical approaches (Biamonte et al., 2017; Devadas & Sowmya, 2025; Lu et al., 2025; Melnikov et al., 2023). These capabilities are especially relevant in smart city environments, where data is often heterogeneous, dynamic, and subject to noise or malicious interference.

One of the most significant contributions of quantum computing to smart infrastructure lies in its ability to solve complex optimization problems. Urban systems inherently involve large-scale, multi-objective optimization challenges, such as traffic flow management, energy distribution, and resource allocation. Quantum optimization algorithms, including the Quantum Approximate Optimization Algorithm (QAOA) and quantum annealing techniques, have demonstrated potential advantages in exploring large solution spaces and identifying near-optimal solutions efficiently (Farhi et al., 2022; Liu et al., 2024; Neukart et al., 2017; Pomeroy et al., 2025; Date et al., 2019; Kim et al., 2025; Heidari et al., 2024; Yulianti et al., 2023; Salloum et al., 2024; Kotsuki et al., 2024). These methods can be applied to optimize traffic signal timing, reduce congestion, improve public transportation scheduling, and enhance the allocation of critical urban resources. By enabling more efficient decision-making, quantum-enhanced optimization can significantly improve the overall performance and sustainability of smart city systems.

In the context of energy systems, quantum-enhanced AI can play a crucial role in optimizing power grid operations and improving system reliability. Modern smart grids are increasingly complex, integrating renewable energy sources such as solar and wind, which introduce variability and uncertainty into the system. Quantum machine learning models can be used to analyze large volumes of sensor data, predict demand patterns, and detect anomalies in real time, thereby enabling more effective grid management (Orka et al., 2025; Gupta et al., 2025; Bukkarayasamudram et al., 2025). Additionally, quantum optimization techniques can be applied to energy dispatch and load balancing problems, ensuring efficient utilization of resources while minimizing operational costs and environmental impact (Liu et al., 2024; Iovane, 2025).

Another important application of quantum-enhanced AI in smart infrastructure is anomaly detection and system monitoring. Urban systems are vulnerable to a wide range of disruptions, including cyberattacks, equipment failures, and environmental events. Quantum-enhanced anomaly detection methods can leverage high-dimensional feature representations and probabilistic modeling to identify unusual patterns in system behavior, enabling early detection of faults or malicious activities (West et al., 2023a; West et al., 2023b; Huang et al., 2023). This capability is essential for maintaining the reliability and security of critical infrastructure systems, particularly in scenarios where rapid response is required to prevent cascading failures.

Hybrid quantum–classical architectures further enhance the applicability of these techniques by combining the scalability of classical systems with the advanced computational capabilities of quantum processors. In such architectures, classical systems handle large-scale data collection and preprocessing, while quantum components perform specialized tasks such as optimization, feature transformation, or probabilistic inference (Benedetti et al., 2019; Cerezo et al., 2021; Abbas et al., 2021). This hybrid approach is particularly relevant in the current era of Noisy Intermediate-Scale Quantum (NISQ) devices, where fully quantum solutions are not yet practical but meaningful advantages can still be achieved through integration with classical AI systems.

Security is another critical concern in smart infrastructure systems, as these systems often manage sensitive data and critical services. Quantum-enhanced AI can contribute to improved security by enabling more robust detection of adversarial activities and supporting the integration of quantum-safe cryptographic techniques (Alexeev et al., 2025; Sen, 20252). For example, quantum-resistant encryption methods can be used to secure communication between devices, while quantum-enhanced anomaly detection systems can identify potential cyber threats in real time. These capabilities are essential for protecting smart city infrastructure against both classical and emerging quantum-enabled attacks.

Future, smart cities may evolve into fully autonomous, self-regulating ecosystems powered by quantum-enhanced AI and advanced communication technologies. The integration of quantum communication networks, such as quantum key distribution (QKD), can provide secure and tamper-resistant data exchange between infrastructure components, ensuring the integrity and confidentiality of critical information (Neilsen & Chuang, 2010; Preskill, 2018; Sen, 2025). Furthermore, the combination of quantum computing, edge AI, and Internet of Things (IoT) technologies may enable real-time, decentralized decision-making across urban systems, allowing cities to adapt dynamically to changing conditions and unforeseen disruptions.

In addition to improving efficiency and security, quantum-enhanced AI may also support sustainability initiatives in smart cities. By optimizing energy usage, reducing traffic congestion, and improving resource allocation, these technologies can contribute to lower carbon emissions and more sustainable urban development. Moreover, the ability to model and simulate complex urban systems using quantum computing may provide new insights into long-term planning and policymaking, enabling more informed decisions that balance economic growth with environmental considerations.

In summary, quantum-enhanced robust AI has the potential to significantly transform smart infrastructure and urban systems by enabling more efficient optimization, improved anomaly detection, and stronger resilience against adversarial threats. While challenges remain in terms of scalability, hardware limitations, and integration complexity, ongoing advancements in quantum computing and AI are likely to drive the development of next-generation smart city systems that are intelligent, secure, and sustainable.

## Defense, National Security, and Strategic Systems

In defense and national security applications, the robustness, reliability, and security of artificial intelligence systems are of paramount importance. AI technologies are increasingly deployed in critical military and strategic domains, including surveillance, intelligence analysis, autonomous systems, cyber defense, and decision support. These systems often operate in highly adversarial environments where malicious actors actively attempt to manipulate data, disrupt communication, or degrade system performance. Adversarial attacks on AI models can compromise surveillance systems, mislead target recognition algorithms, disrupt communication networks, and ultimately lead to incorrect or suboptimal strategic decisions (Biggio & Roli, 2018; Chakraborty et al., 2021; Han et al, 2023; Wendlinger et al., 2024). Such vulnerabilities pose significant risks to national security and highlight the urgent need for robust and secure AI frameworks.

Quantum-enhanced robust AI offers a transformative approach to addressing these challenges by combining the computational advantages of quantum computing with advanced machine learning techniques. One of the key benefits of quantum computing in defense applications lies in its ability to process and analyze large-scale, high-dimensional datasets more efficiently than classical systems. Defense and intelligence operations often involve massive volumes of data, including satellite imagery, radar signals, communication logs, and intelligence reports. Quantum machine learning models can leverage high-dimensional Hilbert spaces to extract meaningful patterns and correlations from such data, enabling faster and more accurate analysis (Biamonte et al., 2017; Devadas & Sowmya, 2025; Lu et al., 2024; Melnikov et al., 2023). These capabilities can significantly enhance situational awareness and support more informed decision-making in complex operational environments.

In addition to data analysis, quantum-enhanced AI can improve the robustness of decision-making systems used in defense applications. Adversarial machine learning has demonstrated that classical AI models can be manipulated through carefully crafted perturbations, potentially leading to catastrophic outcomes in mission-critical scenarios (Goodfellow et al., 2015; Madry et al., 2018; Szegedy, 2014). Quantum-enhanced learning frameworks, including variational quantum circuits and quantum kernel methods, offer alternative approaches to feature representation and optimization that may reduce susceptibility to such attacks (Havlíček et al., 2019; Schuld, M., & Petruccione, 2021; Blank et al., 2020; Cerezo et al., 2021). Recent studies have shown that quantum machine learning models can exhibit improved adversarial robustness due to their probabilistic nature and the complexity of quantum feature spaces (West et al., 2023a; West et al., 2023b; Gong et al., 2024; Dowling et al., 2024; Huang et al., 2023; Nowmi et al., 2025). These properties make it more challenging for adversaries to construct effective attack strategies, thereby enhancing the security of AI-driven defense systems.

Another critical application of quantum-enhanced AI in national security is anomaly detection and threat intelligence. Detecting unusual patterns in data is essential for identifying potential threats, such as cyber intrusions, unauthorized activities, or emerging security risks. Quantum-enhanced anomaly detection techniques can exploit quantum parallelism and probabilistic modeling to analyze complex datasets and identify subtle deviations from normal behavior (West et al., 2023a; West et al., 2023b; Huang et al, 2023). These techniques can be applied to a wide range of defense scenarios, including monitoring communication networks, analyzing sensor data, and detecting anomalies in strategic operations.

Quantum optimization algorithms further contribute to defense applications by enabling efficient planning and resource allocation. Military operations often involve complex optimization problems, such as mission planning, logistics, and resource deployment. Quantum algorithms such as the Quantum Approximate Optimization Algorithm (QAOA) and quantum annealing methods can explore large solution spaces more effectively than classical approaches, potentially providing near-optimal solutions in real time (Farhi et al., 2022; Neukart et al., 2017; Pomeroy et al., 2025; Date et al., 2019; Kim et al., 2025; Heidari et al., 2024; Yulianti et al., 2023; Salloum et al., 2024; Kotsuki et al., 2024; Liu et al., 2024). These capabilities can enhance operational efficiency and improve the effectiveness of strategic decision-making processes.

Secure communication is another critical aspect of defense systems, and quantum technologies offer significant advantages in this domain. Quantum communication protocols, particularly quantum key distribution (QKD), enable the secure exchange of cryptographic keys with provable security guarantees based on the principles of quantum mechanics (Nielsen & Chuang, 2010; Preskill; 2018). Unlike classical encryption methods, which may be vulnerable to quantum attacks, QKD ensures that any attempt at eavesdropping can be detected, thereby providing a higher level of security for sensitive communications. The integration of quantum communication with AI-driven systems can further enhance the security and reliability of defense networks (Alexeev et al., 2025; Sen, 2025).

Hybrid quantum–classical architectures are expected to play a crucial role in the practical deployment of these technologies. Given the current limitations of quantum hardware, such as noise and limited qubit counts, fully quantum solutions remain challenging to implement at scale. Hybrid systems combine classical computational power with quantum-enhanced components, enabling efficient data processing while leveraging quantum advantages for specific tasks such as optimization, pattern recognition, and probabilistic inference (Benedetti et al., 2019; Cerezo et al., 2021; Abbas et al., 2021). These architectures provide a pragmatic pathway for integrating quantum-enhanced AI into existing defense infrastructures.

Defense and national security systems are likely to evolve into highly adaptive and resilient platforms powered by quantum-enhanced AI. These systems may enable real-time situational awareness, autonomous threat detection, and rapid decision-making in complex and uncertain environments. The integration of quantum computing, artificial intelligence, and advanced communication technologies may lead to the development of intelligent defense ecosystems capable of anticipating and responding to threats proactively.

Furthermore, quantum-enhanced AI may facilitate the development of next-generation strategic systems that can operate in contested and adversarial environments. Such systems could include autonomous surveillance platforms, intelligent command-and-control systems, and secure communication networks that leverage quantum cryptography for enhanced protection. The ability to combine robustness, scalability, and security will be critical for maintaining strategic advantages in future defense scenarios.

In conclusion, quantum-enhanced robust AI holds significant potential for transforming defense, national security, and strategic systems. By enabling more efficient data analysis, improving adversarial resilience, enhancing anomaly detection, and supporting secure communication, these technologies can address some of the most critical challenges in modern defense applications. While significant technical and practical challenges remain, ongoing research at the intersection of quantum computing, artificial intelligence, and cybersecurity is expected to play a pivotal role in shaping the future of secure and intelligent defense systems.

## Emerging and Futuristic Applications

Beyond established domains, the convergence of quantum computing and robust artificial intelligence is expected to unlock a wide range of transformative and futuristic applications. These emerging directions extend the capabilities of AI systems beyond current limitations, particularly in environments characterized by uncertainty, high dimensionality, and adversarial risks.

One of the most promising future directions is the development of *self-adaptive AI systems* that can dynamically evolve in response to adversarial environments. Traditional machine learning models are typically trained offline and deployed with fixed parameters, making them vulnerable to evolving attack strategies. In contrast, quantum-enhanced learning frameworks may enable continuous adaptation by leveraging quantum optimization and probabilistic inference mechanisms. Variational quantum algorithms and hybrid quantum–classical models can iteratively update model parameters in real time, allowing systems to respond to adversarial inputs more effectively (Cerezo et al., 2021; Abbas et al., 2021). Recent studies have also highlighted the potential of quantum reinforcement learning and adaptive quantum circuits to enable dynamic learning in uncertain environments (Devadas & Sowmya, 2025; Lu et al., 2024; Liu et al., 2024). Such systems could play a critical role in applications where threat landscapes evolve rapidly, such as cybersecurity, financial markets, and autonomous systems.

Another emerging application lies in *semiconductor and chip design*, where quantum-enhanced AI has the potential to significantly improve modeling efficiency and design optimization. The semiconductor design process involves solving highly complex optimization problems related to circuit layout, power consumption, and performance constraints. Quantum machine learning techniques, particularly those based on quantum optimization and variational circuits, can explore large design spaces more efficiently than classical methods. Recent research has demonstrated that quantum-enhanced models can achieve measurable improvements in design efficiency, with performance gains reported in areas such as layout optimization and circuit simulation (Liu et al., 2024; Iovane, 2025). These advancements suggest that quantum-enhanced AI could play a pivotal role in the next generation of hardware design, enabling faster innovation cycles and more efficient computing architectures.

In the domain of *space exploration and autonomous extraterrestrial systems*, quantum-enhanced robust AI offers unique advantages. Space missions operate in highly uncertain and resource-constrained environments where communication delays and adversarial conditions—such as radiation-induced noise or unexpected system failures—pose significant challenges. AI systems deployed in such settings must operate autonomously while maintaining high levels of reliability. Quantum-enhanced learning models, with their ability to process complex data and adapt to uncertainty, could enable more robust navigation, fault detection, and decision-making in space systems. Additionally, quantum optimization techniques can improve mission planning and resource allocation, enhancing the efficiency of long-duration missions. These capabilities align with broader research efforts exploring the integration of AI and quantum technologies for advanced scientific and engineering applications (Alexeev et al., 2025; Liu et al., 2024).

*Climate modeling and environmental sustainability* represent another critical area where quantum-enhanced AI may have a transformative impact. Climate systems are inherently complex, involving nonlinear interactions across multiple spatial and temporal scales. Classical models often struggle to capture this complexity with sufficient accuracy and efficiency. Quantum machine learning approaches, particularly those leveraging high-dimensional feature representations and probabilistic modeling, may enable more accurate simulation and prediction of climate dynamics. Furthermore, quantum optimization can support sustainable resource management, including energy distribution, water resource allocation, and carbon emission reduction strategies. Recent studies have emphasized the potential of quantum computing to address large-scale environmental challenges by improving computational efficiency and modeling fidelity (Devadas & Sowmya, 2025; Lu et al., 2024; Iovane, 2025).

In *materials science and drug discovery*, quantum-enhanced AI can accelerate the discovery of new materials and chemical compounds. Quantum systems are naturally suited for simulating molecular interactions, which are computationally expensive for classical methods. By integrating quantum simulation with machine learning, researchers can develop models that predict material properties and chemical behaviors with greater accuracy. This capability is particularly important in applications such as battery technology, renewable energy, and pharmaceutical development. Quantum-enhanced AI can also improve robustness in these models, ensuring that predictions remain reliable even in the presence of noisy or adversarial data inputs. Emerging research in quantum chemistry and quantum machine learning highlights the potential for significant advancements in this area (Biamonte et al., 2017; Dunjko & Briegel, 2018; Melnikov et al., 2023).

Another futuristic application involves *secure and trustworthy human–AI interaction systems.* As AI becomes increasingly integrated into sensitive domains such as healthcare, finance, and governance, ensuring trust, privacy, and robustness becomes paramount. Quantum-enhanced AI systems may enable more secure data processing through integration with quantum cryptographic protocols, such as quantum key distribution (QKD). These systems can provide strong guarantees of data integrity and confidentiality, reducing the risk of adversarial manipulation. Additionally, quantum-enhanced models may improve the interpretability and reliability of AI systems by enabling richer probabilistic representations and more robust decision-making processes (West et al., 2023b; Dowling et al., 2024; Nowmi et al, 2025). This could lead to the development of intelligent assistants and decision-support systems that operate securely in high-stakes environments.

The concept of *quantum internet and distributed quantum AI systems* further expands the scope of future applications. A quantum internet would enable the transmission of quantum information across distributed networks, facilitating secure communication and distributed quantum computation. When combined with AI, this could lead to decentralized intelligent systems capable of collaborative learning and decision-making across geographically distributed nodes. Such systems could be particularly valuable in applications such as global financial networks, distributed healthcare systems, and large-scale scientific collaborations. The integration of quantum communication with AI also introduces new possibilities for secure multi-agent systems that are resilient to adversarial interference (Preskill, 2018; Alexeev et al., 2025).

In the context of *advanced manufacturing and Industry 5.0*, quantum-enhanced AI may enable highly adaptive and resilient production systems. These systems can integrate real-time data from sensors, supply chains, and production processes to optimize operations dynamically. Quantum optimization algorithms can improve scheduling, logistics, and resource allocation, while robust AI models ensure reliable performance in the presence of uncertainties and adversarial disruptions. Such capabilities are expected to play a key role in the evolution of intelligent manufacturing systems that prioritize efficiency, sustainability, and resilience (Iovane, 2025).

Finally, the integration of quantum computing with AI may lead to the development of *general-purpose robust intelligent systems* capable of operating across multiple domains. These systems would combine the strengths of quantum computation, such as parallelism and probabilistic modeling, with advanced machine learning techniques to achieve higher levels of adaptability and robustness. While such systems remain largely conceptual at present, ongoing research in Quantum Artificial Intelligence suggests that they may become feasible as quantum hardware matures and scalable algorithms are developed (Devadas & Sowmya, 2025; Lu et al., 2024; Melnikov et al., 2023; Alexeev et al., 2025).

Despite these exciting possibilities, it is important to recognize that many of these applications are still in early stages of development. Significant challenges remain, including hardware limitations, algorithmic scalability, and the integration of quantum and classical systems. Nevertheless, the rapid pace of research in both quantum computing and artificial intelligence suggests that these barriers may be gradually overcome in the coming years.

In summary, emerging and futuristic applications of quantum-enhanced robust AI span a wide range of domains, from adaptive learning systems and semiconductor design to space exploration, climate modeling, and secure human–AI interaction. These applications highlight the transformative potential of combining quantum computing with robust AI methodologies, paving the way for next-generation intelligent systems capable of operating reliably in complex and adversarial environments.

The diverse applications of quantum-enhanced robust AI are summarized in Table 7.

*Table 7. Applications of Quantum-Enhanced Robust AI in Various Domains*

| Domain | Application | Benefits |
|---|---|---|
| Healthcare | Diagnostics, imaging | Secure, accurate predictions |
| Finance | Fraud detection, risk modeling | Real-time robustness |
| Transportation | Autonomous driving | Safety and reliability |
| Cybersecurity | Intrusion detection | Enhanced threat detection |
| Smart Cities | Energy, traffic | Efficient resource use |
| Defense | Surveillance, intelligence | Secure decision-making |

## CHALLENGES, LIMITATIONS, AND OPEN RESEARCH PROBLEMS

The integration of quantum computing with artificial intelligence for enhancing adversarial robustness presents a compelling research frontier. While the potential benefits of Quantum Artificial Intelligence (QAI) are substantial, the field is still in its formative stages and faces numerous technical, theoretical, and practical challenges. These challenges arise from limitations in quantum hardware, the complexity of quantum algorithms, the immature state of quantum machine learning frameworks, and the inherent difficulties of adversarial robustness in AI systems. This section provides a comprehensive discussion of these challenges and outlines key open research problems that must be addressed to realize the full potential of quantum-enhanced robust AI.

### Limitations of Current Quantum Hardware

One of the most significant barriers to the practical deployment of quantum-enhanced AI systems is the limitation of current quantum hardware. Most existing quantum devices fall under the category of NISQ systems, which are characterized by a limited number of qubits, short coherence times, and high error rates (Preskill, 2018; Bharti et al., 2022). These constraints severely restrict the depth and complexity of quantum circuits that can be executed reliably.

Quantum noise and decoherence introduce errors that can accumulate rapidly during computation, leading to unreliable outputs. While *variational quantum algorithms* (VQAs) are designed to mitigate some of these challenges by using shallow circuits and hybrid optimization approaches, their performance is still limited by hardware imperfections (Cerezo et al., 2021). In adversarial settings, where robustness and precision are critical, such noise can further complicate the reliability of quantum-enhanced models.

Scalability is another major concern. Current quantum processors typically support tens to a few hundred qubits, which is insufficient for large-scale machine learning tasks that require processing high-dimensional data. Although ongoing research in quantum error correction and fault-tolerant quantum computing aims to address these limitations, practical large-scale quantum systems are still years away (Bharti et al., 2022).

### Challenges in Quantum Data Encoding

Efficient encoding of classical data into quantum states remains one of the most fundamental challenges in quantum machine learning. While several encoding strategies, such as basis encoding, amplitude encoding, and angle encoding, have been proposed, each comes with trade-offs in terms of computational complexity and representational efficiency (Schuld & Petruccione, 2018; Biamonte et al., 2017).

Amplitude encoding, for instance, offers exponential compression of data but requires complex state preparation procedures that may negate potential quantum advantages. Similarly, angle encoding is easier to implement but may not fully exploit the expressive power of quantum systems. The choice of encoding scheme has a direct impact on model performance, scalability, and robustness.

Recent work on quantum feature maps and embedding strategies has shown promise in improving the representation of data in quantum Hilbert spaces (Havlíček et al., 2019; Schuld & Petruccione, 2021).

However, designing encoding methods that are both efficient and robust to adversarial perturbations remains an open research problem. In adversarial contexts, small perturbations in classical data may translate into complex transformations in quantum states, making robustness analysis more challenging.

## Algorithmic Limitations and Scalability Issues

Although quantum algorithms such as the Quantum Approximate Optimization Algorithm (QAOA) and Variational Quantum Eigensolver (VQE) have demonstrated potential for solving optimization problems, their practical applicability to large-scale machine learning remains limited. Many quantum algorithms rely on assumptions such as sparsity, well-conditioned matrices, or specific data distributions, which may not hold in real-world scenarios (Liu et al., 2024).

Furthermore, the theoretical speedups offered by certain quantum algorithms, such as the Harrow-Hassidim-Lloyd (HHL) algorithm, are often contingent on idealized conditions, including efficient data loading and error-free computation (Arunachalam & de Wolf, 2017). In practice, these assumptions are difficult to satisfy, limiting the applicability of such algorithms in adversarial machine learning contexts.

Another critical issue is the *barren plateau problem,* where gradients in variational quantum circuits vanish exponentially as the system size increases, making training extremely difficult (Cerezo et al., 2021). This problem poses a significant challenge for developing scalable quantum neural networks and hybrid models capable of handling complex datasets.

## Adversarial Robustness in Quantum Machine Learning

While quantum machine learning has been proposed as a potential solution for improving adversarial robustness, recent studies suggest that quantum models are not inherently immune to adversarial attacks. Research has demonstrated that quantum classifiers can also be vulnerable to carefully crafted perturbations, like their classical counterparts (Wendlinger et al., 2024; Nowmi et al., 2025).

However, quantum models may exhibit different robustness characteristics due to their probabilistic nature and high-dimensional feature spaces. For example, quantum kernel methods and variational circuits can create complex decision boundaries that may be more difficult to exploit using traditional attack methods (West et al., 2023). At the same time, new types of quantum-specific adversarial attacks may emerge, targeting quantum state preparation or measurement processes.

Developing a comprehensive theoretical framework for adversarial robustness in quantum machine learning is still an open problem. Existing work on adversarial robustness guarantees for quantum classifiers provides initial insights but is limited in scope (Dowling et al., 2024; Gong et al., 2024). Further research is needed to understand how adversarial perturbations propagate through quantum circuits and how robustness can be formally defined and quantified in quantum settings.

## Integration of Quantum and Classical Systems

Hybrid quantum–classical architectures are currently the most practical approach for implementing quantum-enhanced AI systems. However, integrating quantum and classical components introduces several challenges related to system design, communication overhead, and optimization.

One key issue is the *bottleneck between classical and quantum computation,* particularly during iterative optimization processes. Variational algorithms require repeated evaluations of quantum circuits, with parameters updated by classical optimizers. This iterative loop can be computationally expensive and may limit scalability (Cerezo et al., 2021).

Furthermore, ensuring seamless interoperability between classical and quantum systems requires the development of new software frameworks, programming languages, and hardware interfaces. While platforms such as Qiskit and Cirq have made significant progress, standardized tools for large-scale quantum AI development are still lacking (Alexeev et al., 2025; Lu et al., 2024; Melnikov et al., 2023).

## Security Challenges in the Quantum Era

The integration of quantum computing into AI systems introduces new security challenges that extend beyond adversarial machine learning. Quantum computers have the potential to break widely used cryptographic protocols, necessitating the development of *quantum-resistant cryptography* (Alexeev et al., 2025).

In the context of AI, ensuring end-to-end security requires protecting not only the model but also the data, communication channels, and computational infrastructure. Multi-layered security frameworks that combine quantum-safe cryptography, secure multi-party computation, and blockchain technologies are being explored to address these challenges.

Moreover, the emergence of quantum adversarial attacks raises concerns about the security of quantum machine learning systems themselves. For instance, attackers may exploit vulnerabilities in quantum circuits or manipulate quantum states during computation. Addressing these risks requires a deeper understanding of quantum security principles and the development of robust defense mechanisms.

## Lack of Benchmarking, Standards, and Evaluation Metrics

Another critical challenge in the field is the lack of standardized benchmarks and evaluation metrics for quantum-enhanced adversarial robustness. In classical machine learning, robustness is typically evaluated using well-established benchmarks such as FGSM and PGD attacks. However, similar standardized frameworks do not yet exist for quantum machine learning.

The absence of common evaluation protocols makes it difficult to compare different approaches and assess their effectiveness. Recent efforts to benchmark quantum adversarial robustness have provided valuable insights, but more comprehensive and standardized evaluation frameworks are needed (West et al., 2023; Wendlinger et al., 2024).

Developing such benchmarks will require collaboration across disciplines, including quantum computing, machine learning, and cybersecurity. These benchmarks should account for both classical and quantum attack models, as well as hybrid attack scenarios.

## Interdisciplinary Research Challenges

Quantum-enhanced adversarial robustness lies at the intersection of multiple disciplines, including quantum physics, computer science, machine learning, and cybersecurity. Bridging these fields presents significant challenges, as researchers often have specialized expertise in only one domain.

Effective progress in this area requires interdisciplinary collaboration and the development of educational resources that can bridge knowledge gaps. Training researchers with expertise in both quantum computing and machine learning is essential for advancing the field.

Table 8 outlines key challenges and open research problems in quantum-enhanced adversarial robustness.

*Table 8. Challenges and Open Research Problems in Quantum-Enhanced AI Security*

| Challenge | Description | Research Direction |
|---|---|---|
| Hardware Limitations | Noise, decoherence | Fault-tolerant computing |
| Scalability | Limited qubits | Efficient algorithms |
| Data Encoding | Efficient mapping | Advanced feature maps |
| Robustness Guarantees | Lack of theory | Formal verification |
| Hybrid Integration | System complexity | Architecture design |
| Security | Quantum-safe AI | Cryptographic integration |

## Open Research Problems and Future Directions

Despite the challenges discussed in the preceding sections, the field of quantum-enhanced adversarial robustness remains rich with opportunities for transformative research. The convergence of quantum computing and artificial intelligence has opened new conceptual pathways, but realizing its full potential requires addressing a number of foundational and interdisciplinary research problems. These challenges are not merely incremental extensions of classical machine learning issues; rather, they demand fundamentally new approaches that integrate quantum mechanics, optimization theory, and robust learning paradigms.

One of the most critical areas of future research lies in the development of *robust quantum feature representations*. In classical adversarial machine learning, vulnerabilities often arise due to the way data is represented in feature space. This issue becomes even more complex in quantum systems, where classical data must be encoded into quantum states through various embedding strategies. While techniques such as amplitude encoding, angle encoding, and quantum feature maps have been proposed, their robustness against adversarial perturbations remains insufficiently understood. Since small perturbations in classical input data can lead to nontrivial transformations in quantum Hilbert spaces, designing encoding mechanisms that preserve meaningful structure while minimizing adversarial sensitivity is a significant open problem. Advances in quantum kernel methods and feature space engineering provide promising directions, but a systematic framework for robust quantum representations is still lacking (Havlíček et al., 2019; Schuld & Petruccione, 2021; West et al., 2023).

Closely related to this challenge is the need for *scalable quantum machine learning models*. Current quantum systems are constrained by hardware limitations, including noise, limited qubit counts, and short coherence times, which restrict the size and depth of implementable models. While hybrid quantum–classical architectures have emerged as a practical solution in the NISQ era, their scalability remains an open question. Developing algorithms that can effectively utilize near-term quantum devices while maintaining robustness and efficiency is a major research priority. This includes addressing issues such as barren plateaus in variational quantum circuits, optimizing parameterized models, and improving the stability of quantum training processes (Cerezo et al., 2021; Bharti et al., 2022). Achieving scalability will also require innovations in quantum hardware, error correction techniques, and resource-efficient algorithm design.

Another fundamental research direction involves the establishment of *formal robustness guarantees for quantum machine learning models*. In classical machine learning, significant progress has been made in developing theoretical frameworks for certifying robustness against adversarial attacks. However, similar frameworks for quantum models are still in their infancy. The probabilistic nature of quantum measurement, combined with the complexity of quantum state evolution, makes it challenging to define and quantify robustness in a rigorous manner. Recent work on adversarial robustness guarantees for quantum classifiers has provided initial insights, but a comprehensive theory that generalizes across different quantum architectures and attack models is still needed (Dowling et al., 2024; Gong et al., 2024). Developing such theoretical foundations will be essential for deploying quantum-enhanced AI systems in safety-critical applications.

In parallel, there is a growing need to understand *quantum-aware adversarial attack models*. While much of the current research focuses on defending against classical adversarial attacks, the emergence of quantum computing introduces new possibilities for adversarial strategies. Attackers may exploit quantum properties such as superposition and entanglement to design more sophisticated attacks that are difficult to detect or mitigate using classical methods. Conversely, adversarial techniques may also target vulnerabilities specific to quantum systems, such as errors in state preparation or measurement processes. Understanding how adversarial threats evolve in quantum environments is therefore essential for developing effective defense mechanisms. This requires extending existing adversarial frameworks to incorporate quantum-specific considerations and exploring new attack paradigms that leverage quantum computational capabilities (Wendlinger et al., 2024; Nowmi et al., 2025).

The design of *efficient hybrid quantum–classical architectures* represents another important area of research. Given the limitations of current quantum hardware, hybrid models that combine classical neural networks with parameterized quantum circuits offer a practical pathway for leveraging quantum advantages. However, optimizing these architectures presents several challenges, including the efficient integration of classical and quantum components, minimizing communication overhead, and ensuring stable training dynamics. Developing standardized frameworks for hybrid learning, along with efficient optimization algorithms, will be crucial for advancing this field. Furthermore, understanding how to allocate computational tasks between classical and quantum components in a way that maximizes performance and robustness remains an open problem (Abbas et al., 2021; Cerezo et al., 2021).

Security considerations also give rise to the need for *quantum-secure AI frameworks* that integrate robust machine learning with quantum-safe cryptographic techniques. As quantum computing threatens to undermine classical cryptographic systems, ensuring the security of AI models and data pipelines becomes increasingly important. Future research must explore how quantum-resistant encryption methods, secure multi-party computation, and blockchain technologies can be combined with robust AI systems to provide end-to-end security. At the same time, it is necessary to address potential vulnerabilities introduced by quantum computing itself, including quantum adversarial attacks and side-channel threats. This calls for a holistic approach to security that spans both classical and quantum domains (Alexeev et al., 2025; Devadas & Sowmya, 2025).

Another important direction involves the development of *application-specific solutions* that tailor quantum-enhanced robustness techniques to specific domains. While general-purpose frameworks are valuable, different application areas, such as healthcare, finance, autonomous systems, and cybersecurity, have unique requirements and constraints. For instance, healthcare applications demand high levels of interpretability and reliability, while financial systems require real-time decision-making and regulatory compliance. Designing quantum-enhanced AI models that address these domain-specific challenges will be essential for practical deployment. Recent studies in quantum machine learning for healthcare and industrial applications highlight the importance of domain-aware approaches in achieving meaningful performance gains (Gupta et al., 2025; Bukkarayasamudram et al., 2025).

Finally, achieving practical quantum advantage in adversarial machine learning will require *coordinated interdisciplinary research efforts*. The integration of quantum computing and AI spans multiple domains, including physics, computer science, mathematics, and engineering. Bridging these disciplines is essential for addressing the complex challenges outlined above. Recent surveys have emphasized that progress in quantum machine learning depends not only on advances in algorithms and hardware but also on the development of new theoretical frameworks and collaborative research ecosystems (Devadas & Sowmya, 2025; Lu et al., 2024; Melnikov et al., 2023). Building such ecosystems will involve fostering collaboration between academia, industry, and government institutions, as well as developing educational programs that equip researchers with the necessary interdisciplinary skills.

The future of quantum-enhanced adversarial robustness is defined by a set of deeply interconnected research challenges that span representation learning, algorithm design, theoretical analysis, system integration, and security. Addressing these challenges will require a combination of theoretical innovation, experimental validation, and technological advancement. While significant obstacles remain, the rapid pace of progress in both quantum computing and artificial intelligence suggests that these challenges are not insurmountable. Continued research in this area has the potential to redefine the foundations of secure and robust AI, paving the way for intelligent systems that can operate reliably in increasingly complex and adversarial environments.

Figure 11 summarizes the key challenges and future research directions.

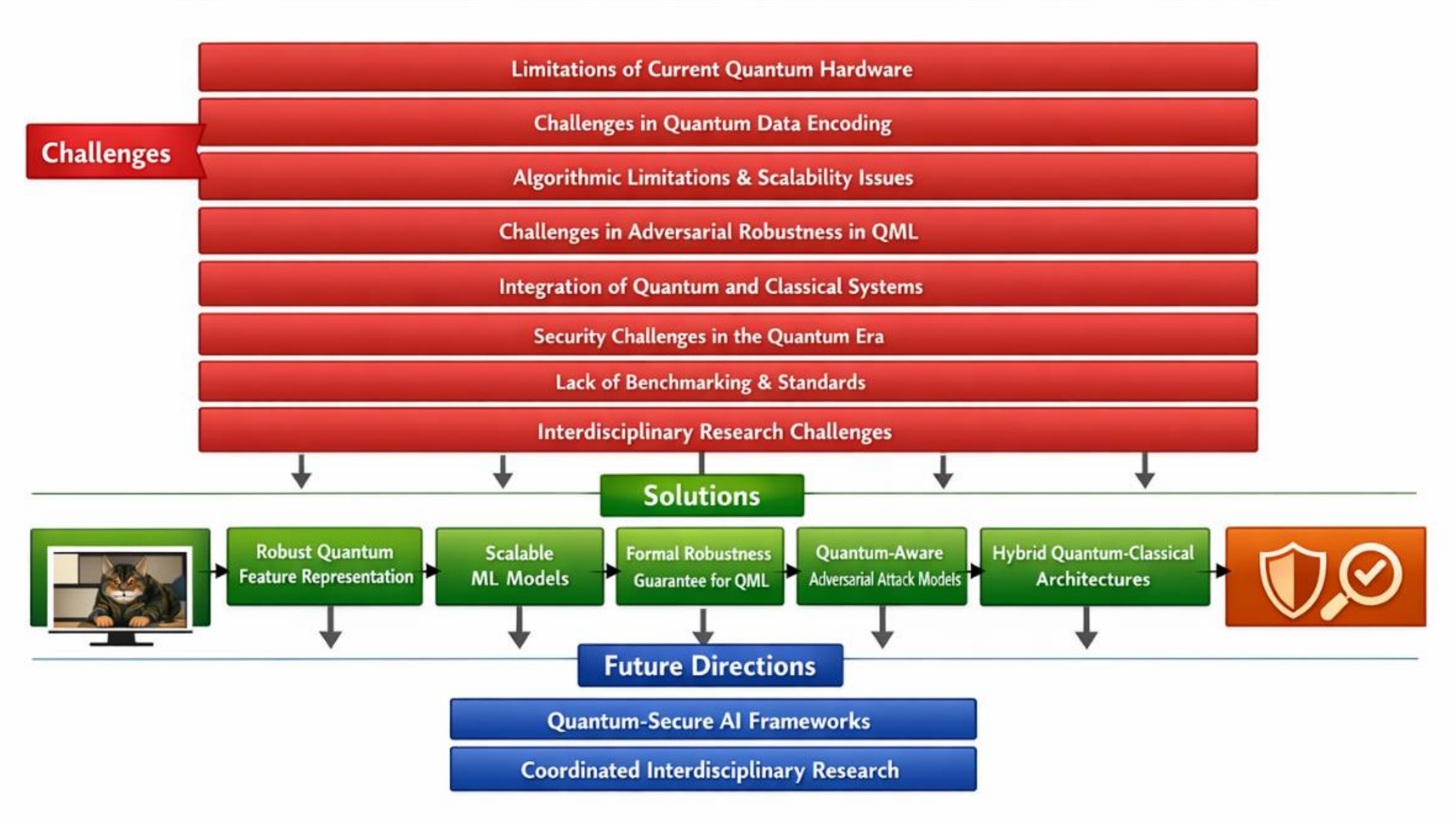

*Figure 11. Research challenges and Future Directions in Quantum-Enhanced AI Security*

## CONCLUSION

The rapid evolution of artificial intelligence has fundamentally transformed modern digital ecosystems, enabling unprecedented capabilities in data analysis, automation, and intelligent decision-making. However, as AI systems become increasingly embedded in critical applications, their vulnerabilities, particularly to adversarial attacks, have emerged as a major concern. Adversarial machine learning has revealed that even highly accurate models can be manipulated through carefully crafted perturbations, raising serious questions about the reliability, security, and trustworthiness of AI-driven systems (Goodfellow et al., 2015; Madry et al., 2018; Chakraborty et al., 2021). Addressing these challenges requires not only incremental improvements in existing techniques but also fundamentally new paradigms for designing robust and secure intelligent systems.

In parallel, quantum computing has emerged as a transformative computational paradigm with the potential to revolutionize how complex problems are solved. By leveraging the principles of superposition, entanglement, and quantum interference, quantum systems can explore high-dimensional solution spaces in ways that are not feasible for classical computers (Nielsen & Chuang, 2010; Preskill, 2018). Although still in its early stages of development, quantum computing has already demonstrated promising capabilities in areas such as optimization, simulation, and machine learning, laying the foundation for the emerging field of Quantum Artificial Intelligence (QAI) (Biamonte et al., 2017; Dunjko & Briegel, 2018; Devadas & Sowmya, 2025). The integration of quantum computational techniques with artificial intelligence introduces new opportunities to address some of the most pressing challenges in AI, particularly those related to scalability, efficiency, and robustness.

This chapter has explored the intersection of adversarial machine learning and quantum computing, presenting a comprehensive perspective on how quantum-enhanced approaches can contribute to the development of secure and reliable AI systems. Beginning with an overview of adversarial attacks and their implications, the discussion highlighted the inherent vulnerabilities of modern machine learning models and the limitations of existing defense mechanisms. Techniques such as adversarial training, defensive distillation, and robust optimization have provided partial solutions, but they often fall short in the face of adaptive and sophisticated attack strategies (Papernot et al., 2016; Biggio & Roli, 2018). These limitations underscore the need for innovative approaches that go beyond classical computational paradigms.

The introduction of quantum computing concepts in this chapter provided a foundational understanding of how quantum systems differ from classical ones and why they hold promise for advancing AI. Concepts such as qubits, quantum gates, and quantum circuits form the building blocks of quantum algorithms, enabling new forms of computation that can potentially enhance learning processes. The discussion further extended to quantum machine learning models, including variational quantum circuits, quantum kernel methods, and hybrid quantum–classical architectures, which represent practical approaches for leveraging quantum capabilities in the current NISQ era (Cerezo et al., 2021; Schuld & Petruccione, 2021).

A central theme of this chapter has been the exploration of conceptual frameworks for *quantum-enhanced adversarial robustness*. Quantum optimization algorithms, such as the Quantum Approximate Optimization Algorithm (QAOA), offer new possibilities for improving adversarial training by enabling more efficient exploration of complex loss landscapes (Farhi et al., 2022). Similarly, quantum feature representations can map data into high-dimensional Hilbert spaces, potentially improving class separability and reducing susceptibility to adversarial perturbations (Havlíček et al., 2019). Hybrid architectures that integrate classical neural networks with parameterized quantum circuits further provide a flexible framework for combining the strengths of both paradigms, enabling improved robustness and computational efficiency (Abbas et al., 2021).

The application domains discussed in this chapter illustrate the far-reaching implications of quantum-enhanced robust AI. From healthcare and finance to autonomous systems and cybersecurity, the ability to develop AI models that are resilient to adversarial manipulation is critical for ensuring system integrity and user trust. Quantum-enhanced approaches have the potential to improve not only the performance of these systems but also their reliability in adversarial environments. At the same time, emerging applications in areas such as smart infrastructure, defense systems, and advanced manufacturing highlight the transformative potential of integrating quantum computing with robust AI methodologies (Alexeev et al., 2025; Lu et al., 2024).

Despite these promising developments, the chapter has also emphasized the significant challenges that must be addressed before quantum-enhanced adversarial robustness can be realized in practice. Limitations in quantum hardware, including noise, decoherence, and scalability constraints, remain major obstacles (Preskill, 2018; Bharti et al., 2022). Algorithmic challenges, such as the barren plateau problem in variational circuits and the complexity of quantum data encoding, further complicate the development of effective quantum machine learning models. Additionally, the theoretical foundations of adversarial robustness in quantum systems are still evolving, and a comprehensive understanding of quantum-specific attack and defense mechanisms is yet to be established (Dowling et al., 2024; Nowmi et al., 2025).

The future of quantum-enhanced adversarial robustness will depend on sustained interdisciplinary collaboration and continued advancements in both quantum computing and artificial intelligence. Progress in quantum hardware, including the development of fault-tolerant systems, will be essential for scaling quantum machine learning models to real-world applications. At the same time, advances in algorithm design, optimization techniques, and hybrid architectures will play a critical role in bridging the gap between theoretical potential and practical implementation.

Furthermore, the development of standardized benchmarks, evaluation frameworks, and robustness metrics will be crucial for assessing the effectiveness of quantum-enhanced defense mechanisms. As the field matures, it is also important to consider the broader ethical, societal, and regulatory implications of deploying quantum-enhanced AI systems, particularly in high-stakes domains where trust and accountability are paramount.

A future roadmap for quantum-enhanced secure AI systems is presented in Figure 12. It illustrates a progressive roadmap for quantum-enhanced secure AI systems, beginning with the current era of NISQ devices characterized by limited scalability and basic quantum algorithms. It transitions into the near-term hybrid phase, where quantum and classical systems are integrated to enhance optimization and cryptographic capabilities. The next stage showcases fault-tolerant quantum AI, emphasizing robust error correction and scalable quantum machine learning models. Finally, the roadmap culminates in fully secure AI ecosystems, featuring autonomous AI agents and global protection mechanisms enabled by end-to-end quantum security.

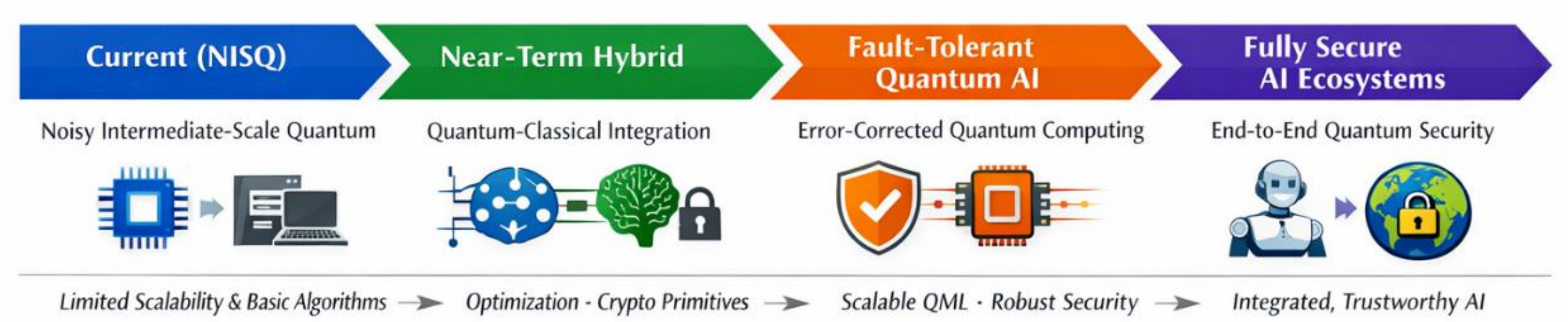

*Figure 12. Evolutionary Timeline of Quantum-Enhanced Secure AI Systems*

In conclusion, the integration of quantum computing with adversarial machine learning represents a promising and rapidly evolving research frontier. While significant challenges remain, the potential benefits of quantum-enhanced robust AI are substantial, offering new pathways for developing intelligent systems that are not only powerful but also secure and trustworthy. As research in this area continues to advance, Quantum Artificial Intelligence may play a pivotal role in shaping the next generation of AI technologies, enabling systems that can operate reliably in increasingly complex and adversarial environments.

## KEY TERMS AND DEFINITIONS

**Artificial Intelligence (AI):** Artificial Intelligence refers to the development of computational systems capable of performing tasks that typically require human intelligence, including learning, reasoning, problem-solving, perception, and decision-making.

**Adversarial Machine Learning:** Adversarial Machine Learning is a subfield of AI that studies the vulnerabilities of machine learning models to maliciously crafted inputs and develops techniques to both exploit and defend against such attacks.

**Adversarial Attack:** An adversarial attack is a technique in which carefully designed perturbations are applied to input data with the intent of causing a machine learning model to produce incorrect or misleading outputs.

**Adversarial Robustness:** Adversarial robustness refers to the ability of a machine learning model to maintain reliable and accurate performance in the presence of adversarial perturbations or malicious inputs.

**Fast Gradient Sign Method (FGSM):** FGSM is a gradient-based adversarial attack technique that generates perturbations by adjusting input data in the direction of the gradient of the loss function, scaled by a small factor.

**Projected Gradient Descent (PGD):** PGD is an iterative adversarial attack method that repeatedly applies small gradient-based perturbations while constraining the modified input within a specified norm bound.

**Adversarial Training:** Adversarial training is a defense mechanism that improves model robustness by incorporating adversarial examples into the training process, enabling the model to learn more resilient representations.

**Quantum Computing:** Quantum computing is a computational paradigm that leverages principles of quantum mechanics, such as superposition and entanglement, to perform information processing tasks beyond the capabilities of classical systems.

**Qubits:** A qubit (quantum bit) is the fundamental unit of quantum information, capable of existing in a superposition of multiple states simultaneously, unlike a classical bit which is restricted to either 0 or 1.

**Quantum Superposition:** Quantum superposition is the property that allows a quantum system to exist in multiple states at the same time, enabling parallel computation across a range of possible configurations.

**Quantum Entanglement:** Quantum entanglement is a phenomenon in which two or more qubits become correlated in such a way that the state of one qubit cannot be described independently of the others, regardless of spatial separation.

**Quantum Machine Learning (QML):** Quantum Machine Learning is an interdisciplinary field that integrates quantum computing techniques with machine learning algorithms to enhance computational efficiency, scalability, and learning performance.

**Variational Quantum Algorithms (VAQs):** Variational Quantum Algorithms are hybrid quantum–classical algorithms that use parameterized quantum circuits optimized through classical feedback loops, making them suitable for noisy intermediate-scale quantum devices.

**Quantum Feature Mapping:** Quantum feature mapping is a technique that encodes classical data into quantum states, often transforming data into high-dimensional Hilbert spaces to improve separability and learning performance.

**Hybrid Quantum-Classical Architecture:** A hybrid quantum–classical architecture combines classical computing components with quantum circuits, enabling practical implementation of quantum machine learning models in current hardware environments.